\documentclass[journal,twoside,web]{ieeecolor2}
\usepackage{generic}
\usepackage{cite}
\usepackage{amsmath,amssymb,amsfonts}
\usepackage{algorithmic}
\usepackage{graphicx}
\usepackage{textcomp}
\usepackage{graphicx}
\usepackage{subcaption}
\usepackage{cite}      
\usepackage{booktabs}
\usepackage{multirow}
\usepackage{makecell}
\usepackage{fancyhdr}
\usepackage{xcolor}
\usepackage{etoolbox}
\usepackage{xparse}
\usepackage{eso-pic}

\makeatletter
\newcommand{\bluebiblist}{}

\let\oldbibitem\bibitem
\RenewDocumentCommand{\bibitem}{om}{%
  \ifinlist{#2}{\bluebiblist}{\color{blue}}{\color{black}}%
  \IfNoValueTF{#1}
    {\oldbibitem{#2}}
    {\oldbibitem[#1]{#2}}%
}
\makeatother

\makeatletter
\let\NAT@parse\undefined  
\makeatother
\usepackage[
    colorlinks=true,
    linkcolor=blue,
    citecolor=blue,
    urlcolor=blue
]{hyperref}

\def\BibTeX{{\rm B\kern-.05em{\sc i\kern-.025em b}\kern-.08em
    T\kern-.1667em\lower.7ex\hbox{E}\kern-.125emX}}
\begin{document}

\title{Advancing Electrolaryngeal Speech Enhancement Through Speech--Text Representation Learning}
\author{Ding Ma, Jinyi Mi, Fengji Li, \IEEEmembership{Member, IEEE}, Lester Phillip Violeta, Jiajun He, Wenchin Huang, \IEEEmembership{Member, IEEE}, Kazuhiro Kobayashi, \IEEEmembership{Member, IEEE}, and Tomoki Toda, \IEEEmembership{Senior Member, IEEE}
\thanks{Manuscript received 10 December 2025. This work was supported in part by JSPS KAKENHI under Grant 25K02788, and in part by the BRIDGE Program (R7-H05), implemented by the Cabinet Office, Government of Japan.
\textit{(Corresponding author: Ding Ma.)}}
\thanks{This work involved human subjects in its research. The collection and use of the data were approved by the Ethics Committee of Nagoya University Hospital under Approval No. 2021-0232.}
\thanks{Ding Ma, Jinyi Mi, Lester Phillip Violeta, Jiajun He, and Wenchin Huang are with the Graduate School of Informatics, Nagoya University, Nagoya 464-8601, Japan (e-mail: ding.ma@g.sp.m.is.nagoya-u.ac.jp).}
\thanks{Fengji Li is  with the School of Biological Science and Medical Engineering, Beihang University, Beijing 100191, China.}
\thanks{Kazuhiro Kobayashi is with the Graduate School of Informatics, Nagoya University, Nagoya 464-8601, Japan, and also with TARVO, Inc., Nagoya 460-0008, Japan.} 
\thanks{Tomoki Toda is with Information Technology Center, Nagoya University, Nagoya 464-8601, Japan.}}

\maketitle

\AddToShipoutPictureFG*{%
  \AtPageLowerLeft{%
    \hspace*{\dimexpr1in+\oddsidemargin\relax}%
    \raisebox{0.35in}[0pt][0pt]{%
      \parbox{\textwidth}{\footnotesize
        \copyright~2026 IEEE. Personal use of this material is permitted.
        This is the accepted version of an article published in
        \textit{IEEE Transactions on Biomedical Engineering}.
        DOI: 10.1109/TBME.2026.3694703.
      }%
    }%
  }%
}

\begin{abstract}
\textit{Objective:} laryngectomees depend on an electromechanical device to generate electrolaryngeal (EL) speech for verbal communication. Compared with normal speech, EL speech suffers from severe distortion, limited phonetic variation, unnatural prosody, and temporal shifts, degrading naturalness and intelligibility. Although sequence-to-sequence (seq2seq) voice conversion (VC) based EL-speech-to-normal-speech conversion (EL2SP) is promising, substantial mismatches between EL and normal speech inevitably cause cumulative mapping errors that limit performance. To address this, we describe a novel representation learning framework integrating speech and text representations to improve mapping and reconstruction quality within a seq2seq VC model. \textit{Methods:} our methodology comprises two main stages: 1) representation integration and learning, and 2) reconstruction training. A network capable of incorporating auxiliary text information is first constructed with pretrained modules to learn speech--text-based integrated representations. Then, an autoencoder-style reconstruction strategy finalizes EL2SP model to inherit these representations without increasing model complexity. Additional optimization designs are performed across these stages. We introduce three fusion strategies including middle-, input-, and hybrid-level fusion strategies that progressively enhance learning. Moreover, besides standard seq2seq VC objectives, an additional reconstruction loss on the integrated representation is introduced to refine representation transfer. \textit{Results:} experiments under different EL2SP datasets consistently demonstrate that our methods, combined with data augmentations, outperform baselines relying solely on speech representations regarding both conversion quality and intelligibility. Furthermore, progressive improvements with system design depth validate the effectiveness of our methods. \textit{Significance:} the proposed methods provide an extensible and practical methodology for EL speech enhancement and assistive communication technologies.
\end{abstract}

\begin{IEEEkeywords}
electrolaryngeal speech, sequence-to-sequence voice conversion, EL-speech-to-normal-speech conversion, speech--text representation learning
\end{IEEEkeywords}

\section{Introduction}
\label{sec:introduction}
\IEEEPARstart{I}{n} human communications, speech conveys a wide range of information that makes interaction natural and effective. The vocal folds are the most crucial organ in phonation. They vibrate the airflow from trachea to produce a source excitation sound. This excitation is modulated by the articulatory configurations of the vocal tract, along with its resonance characteristics, to produce speech~\cite{kobayashi2018electrolaryngeal}. Speech signals from a healthy speaker embody three essential aspects: 1) linguistic content, which transmits explicit meaning through words and grammar; 2) nonlinguistic information, such as speaker identity; and 3) paralinguistic information~\cite{wani2021comprehensive}, including accent, prosody, and emotion.

\subsection{Laryngectomees and Electrolaryngeal Speech}
Individuals who experience damage to or lost control for vocal folds cannot fully convey linguistic, nonlinguistic, and paralinguistic information, resulting in severe communication barriers and life distresses~\cite{babin2009psychosocial,polat2015effects}. A representative case involves patients who undergo total laryngectomy, a surgical removal of the larynx, commonly performed as a treatment for laryngeal cancer~\cite{tang2015voice}. This surgery leads to the complete loss of the original sound source organs including vocal folds, preventing patients from producing speech altogether.  

\begin{figure}[!t]
\centerline{\includegraphics[width=0.80\columnwidth]{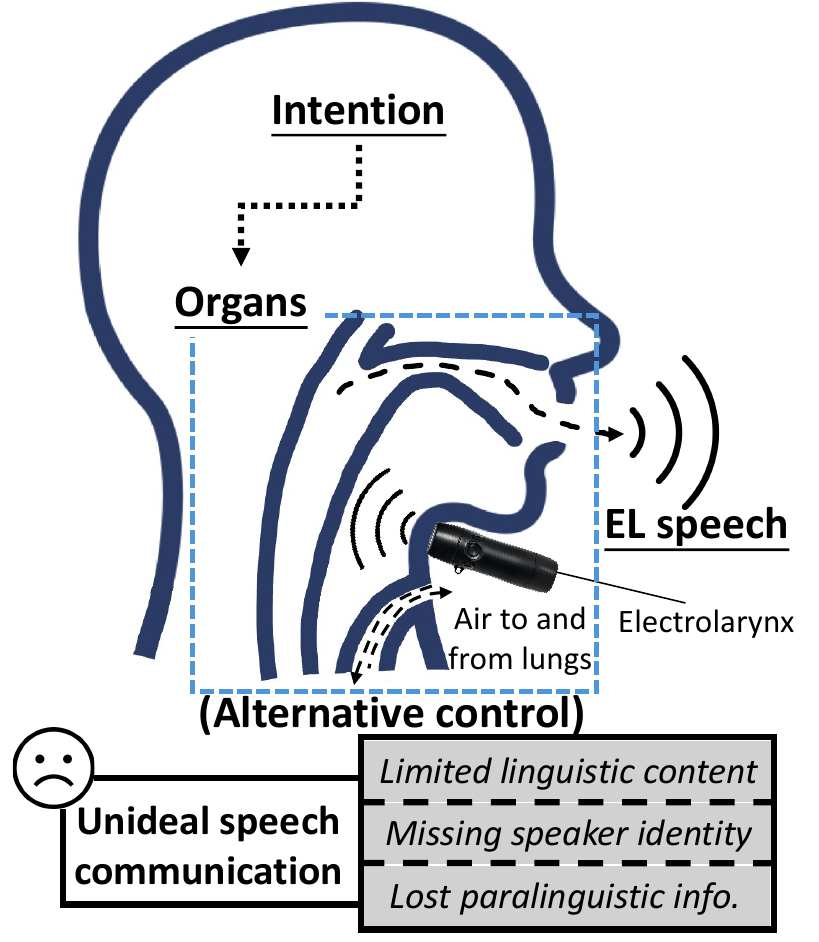}}
\caption{Illustration of EL speech production process and its communication limitations.}
\label{fig1}
\end{figure}

Since most laryngectomees retain the knowledge of speaking and the vocal tract remains intact in clinical practice~\cite{li2024end}, several alternative methods are developed to compensate for the loss of vocal folds and rehabilitate the speaking ability. One is esophageal speech. Patients push or swallow air into the esophagus as a vibration source to produce sound. However, esophageal speech has fundamental constraints: 1) it necessitates months of training, while the effectiveness depends on the extent of laryngectomy, not all the patients can master it; and 2) it typically results in short, discontinuous utterances due to insufficient air supply. Another is tracheoesophageal speech, involving a tracheoesophageal puncture (TEP) surgery to implant a one-way valve that redirects pulmonary air into the esophagus. TEP surgery yields more fluent speech than esophageal speech, but requires additional surgery, continuous medical care, imposing further burdens on patients. In contrast, electrolaryngeal (EL) speech~\cite{singer1980endoscopic,hashiba2001industrialization,williams1985differences} offers a faster and safer option. EL speech is an artificial speech generated by a portable device named electrolarynx, which is placed against the neck to externally simulate the vibrations of the vocal folds. The generated excitation sounds are conducted into the oral cavity and articulated to form EL speech. Many laryngectomees are able to use electrolarynx to produce EL speech within only a few days after surgery. Therefore, EL speech has become a convenient and widely adopted means for patients. However, due to the inherent limitations of its sound generation mechanism, EL speech lacks linguistic clarity and shows degraded paralinguistic expressiveness and speaker identity, as noted in Fig.~\ref{fig1}.

\begin{figure}[!t]
\centerline{\includegraphics[width=\columnwidth]{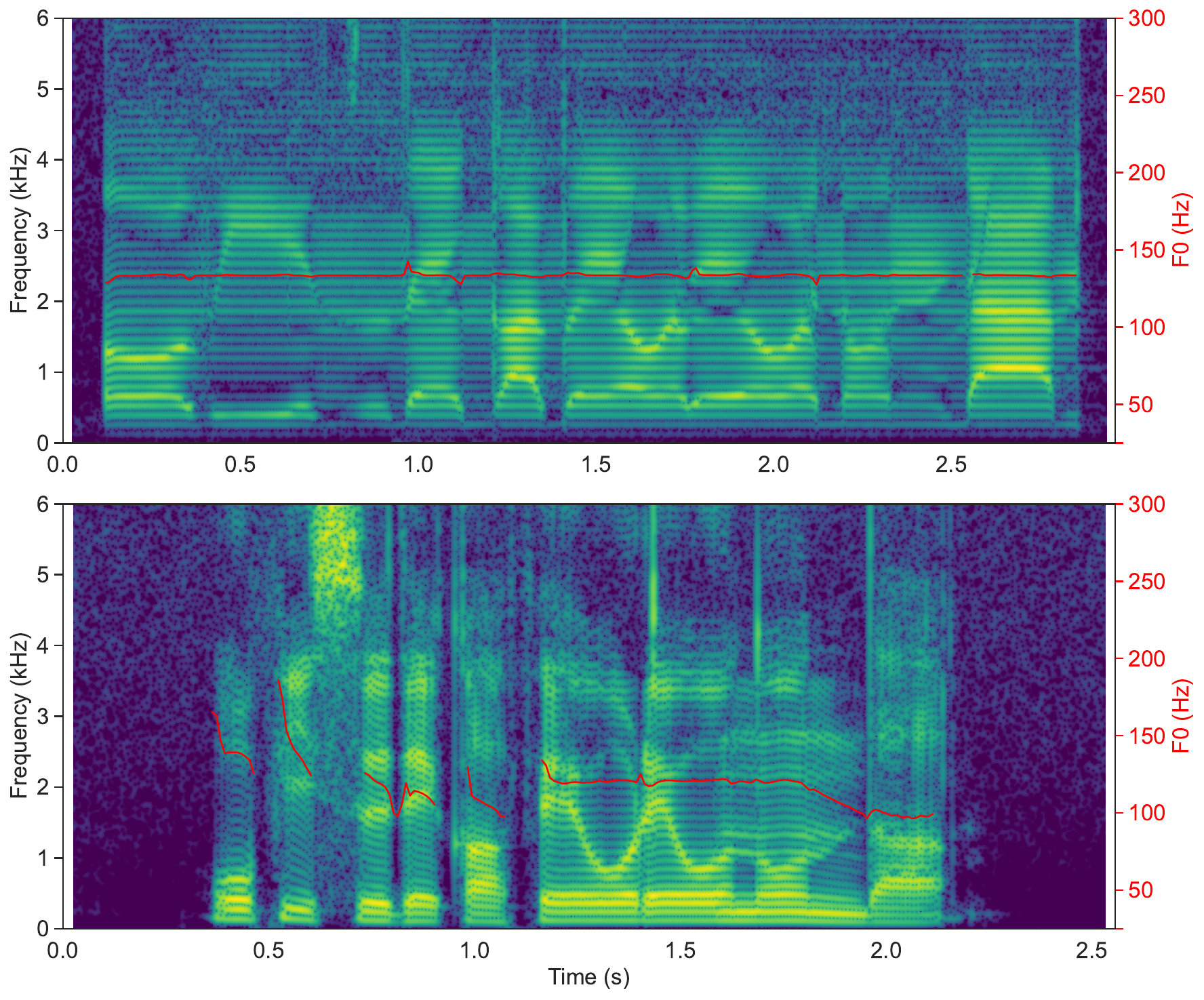}}
\caption{Comparison of mel-spectrograms and their F0 contours between EL speech (upper) and normal speech (lower) for the same utterance.}
\label{fig2}
\end{figure}

Fig.~\ref{fig2} illustrates the gap between EL and normal speech, which primarily stems from two factors. First, the mechanical excitation signal fails to reproduce the natural fundamental frequency (F0) contour in normal voices~\cite{ma1999improvement}, including pitch movements, intonation patterns, and the distinction between voiced and unvoiced segments~\cite{stanislav2017recognition}. Consequently, the natural expressions like prosodic characteristics, and phonemic clarity are largely diminished. Second, the high-intensity excitation inevitably radiates out, introducing strong mechanical noise that degrades both the quality and intelligibility of EL speech. To compensate, EL speakers often adopt an atypical speaking style, e.g., speak at a slower rate, making the temporal structure, including speech duration, to deviate from that of normal speech, as demonstrated in Fig.~\ref{fig2}. Collectively, these deficiencies create substantial mismatches between feature distributions of EL and normal speech, rendering EL speech unnatural, unclear, and unpleasant for both users and listeners, and thus must be addressed to improve patients’ quality of lives.

\subsection{Seq2seq Voice Conversion for EL Speech Enhancement and Remaining Challenge}

Voice conversion (VC) is a methodology originally developed to convert one speaker’s voice into another’s without changing the linguistic content~\cite{childers1985voice,stylianou1998continuous,toda2007voice}. With progress in the field, VC has played a role for EL speech enhancement, namely, EL-speech-to-normal-speech conversion (EL2SP). In VC, the key lies in accurate mapping between source and target features through alignment learning. Common approaches to EL2SP apply conventional statistical VC models, whose performance is determined by rigorous frame-by-frame alignment paradigm~\cite{kobayashi2018electrolaryngeal,nakamura2012speaking,tanaka2014hybrid,doi2013alaryngeal,qian2021mandarin}. It imposes the temporal structure of the converted speech to be identical to that of EL speech, which contradicts the intrinsic differences between EL and normal speech, thereby limiting conversion performance. 

Over the past decade, the emergence of deep neural networks has led to the development of sequence-to-sequence (seq2seq) models~\cite{sutskever2014sequence} to address these limitations. With a sequence-wise encoder-decoder framework, the seq2seq model encodes the entire source sequence into linguistically oriented intermediate representations, which are then decoded into target sequence of potentially different length. Therefore, seq2seq VC can learn complex mappings, enabling it to flexibly adjust the duration of converted speech and achieve better temporal alignment with normal speech. This capability has already demonstrated seq2seq VC is more effective in normal-to-normal VC tasks~\cite{miyoshi2017voice,tanaka2019atts2s}, and it therefore holds strong potential for EL2SP.

One concern with seq2seq VC is its reliance on large-scale, high-quality parallel data, whereas EL2SP data are practically minimal due to strict recording protocols and the inherently fragile physiological conditions of laryngectomees~\cite{11095635,10533680}. A popular solution to maintain the superiority of seq2seq VC is transfer learning using a pretraining--fine-tuning paradigm. Yen et al.~\cite{yen2021mandarin} developed an EL2SP system by first pretraining the model on a large-scale normal corpus and subsequently fine-tuning it to a small-scale EL2SP dataset, resulting in outperformed performance over conventional ones. Following this, Ma et al.~\cite{11095635} further incorporated imperfect synthetic data (SD) as data augmentation to establish a two-stage fine-tuning strategy, achieving state-of-the-art performance.

Nonetheless, these methods encounter performance bottlenecks in EL2SP, as the substantial mismatches between EL and normal speech still hinder effective representation learning. When processing EL speech features, the encoder struggles to extract pure and stable intermediate representations, inevitably introducing cumulative errors. These errors are subsequently propagated to the decoder, which degrades the accuracy and consistency of the overall mapping. Consequently, there is still much room to reach normal-human-level conversion naturalness and intelligibility.

\subsection{Research Motivation and Contributions of our Work}
With all these said, we aim to propose more advanced representation learning method under seq2seq modeling for EL2SP. To promote accurate encoding, a promising direction is to incorporate linguistic information during training. Unlike speech features, linguistic cues from text are inherently speaker-independent and provide a clearer, more structured representation of underlying content. To this end, leveraging text-to-speech (TTS) models represents a natural choice. From the perspective of task formulation, both VC and TTS can be viewed as forms of speech synthesis, and in particular, the encoder in TTS can extract purer linguistic-rich representations than those captured by speech-based VC encoders, enabling higher-quality mappings toward target features. Previous works~\cite{huang2019voice,huang2021pretraining} provided some empirical evidence that transferring pretrained TTS parameters or attention patterns can assist VC. 

In addition, some methods incorporate TTS modules as functional components within the VC training pipeline. Park et al.~\cite{park2020cotatron} introduced a TTS encoder to align speech and text representations, guiding the extraction of more accurate representations for better conversion performance. Likewise, Zhang et al.~\cite{zhang2019joint} proposed a dual-attention framework to jointly train the TTS and VC encoders, so that TTS and VC tasks can mutually reinforce each other. Despite this, these approaches present inherent limitations. First, they increase the architectural complexity of the original VC framework. Specifically, the system built in~\cite{park2020cotatron} requires text as an auxiliary input during inference, while the method in~\cite{zhang2019joint} has a high-level demands on model size, tuned hyperparameters, and training algorithms to ensure its generalized multi-task attention. Moreover, none of these studies have been verified under the more challenging EL2SP scenario.

On the other hand, the aforementioned studies follow indirect forms of linguistic guidance, in which text information is leveraged through TTS modules either as pretrained priors or as alignment references for speech features. Such indirectness prevents full exploitation of text advantages, leaving constrained conversion performance under severe acoustic mismatches. 

A more direct line of research falls into developing multimodal approaches, where additional modalities (e.g., texts or images) are explicitly integrated into speech representations. Such approaches have shown preliminary progress in speech synthesis tasks such as TTS~\cite{chen2024stylefusion,guan2024mm} and VC~\cite{kameoka2019crossmodal,niu2024hybridvc,li2023alignsts}. Here we focus on the latter. Kameoka et al.~\cite{kameoka2019crossmodal} pioneered crossmodal VC by jointly modeling speech and facial information, realizing cross-modal conversion between auditory and visual domains. Additionally, Niu et al.~\cite{niu2024hybridvc} combined text and audio prompts, aligning style embeddings in latent space via contrastive learning for flexible speech style conversion. Furthermore, Li et al.~\cite{li2023alignsts} introduced rhythm cues as an independent modality in speech-to-singing conversion, where melody information was fused with speech via cross-modal alignment to enhance both quality and naturalness. While these studies show the benefits of integrating auxiliary information, they mainly focus on style modeling and depend on crafted prompts or cues as conditioning signals during run-time. Hence, such paradigms are still misaligned with EL2SP scenario, where the system operates single EL input for content recovery rather than style manipulation.

Beyond speech synthesis, some recent studies in broader speech processing tasks have also explored multimodal approaches. For instance, in speech emotion recognition, Li~et al.~\cite{li2025wavfusion} proposed WavFusion, which combines RoBERTa~\cite{zhuang-etal-2021-robustly} and wav2vec~2.0~\cite{baevski2020wav2vec} with a gated cross-modal attention mechanism for text--audio multimodal fusion. However, its performance advantage relies on multimodal inputs, making it less suitable for EL2SP with a single EL input in most practical scenarios. In noisy speech enhancement, Lv et al.~\cite{10994671} proposed HAV-DF to integrate audio--visual modalities for healthcare speech communication, whereas its visual processing pipeline and adaptive-weighted losses increase the complexity of training configuration and architecture. In automatic speech recognition (ASR), Cappellazzo~et al.~\cite{cappellazzo2025large} proposed Llama-AVSR, which adapts a pretrained large language model (LLM) with LoRA~\cite{hu2022lora} modules to enhance ASR and audio--visual speech recognition. Nevertheless, this framework necessitates hundreds of hours of audio--visual data for fine-tuning, which is difficult to satisfy in practical EL2SP scenarios with scarce data.

\begin{figure}[!t]
\centerline{\includegraphics[width=\columnwidth]{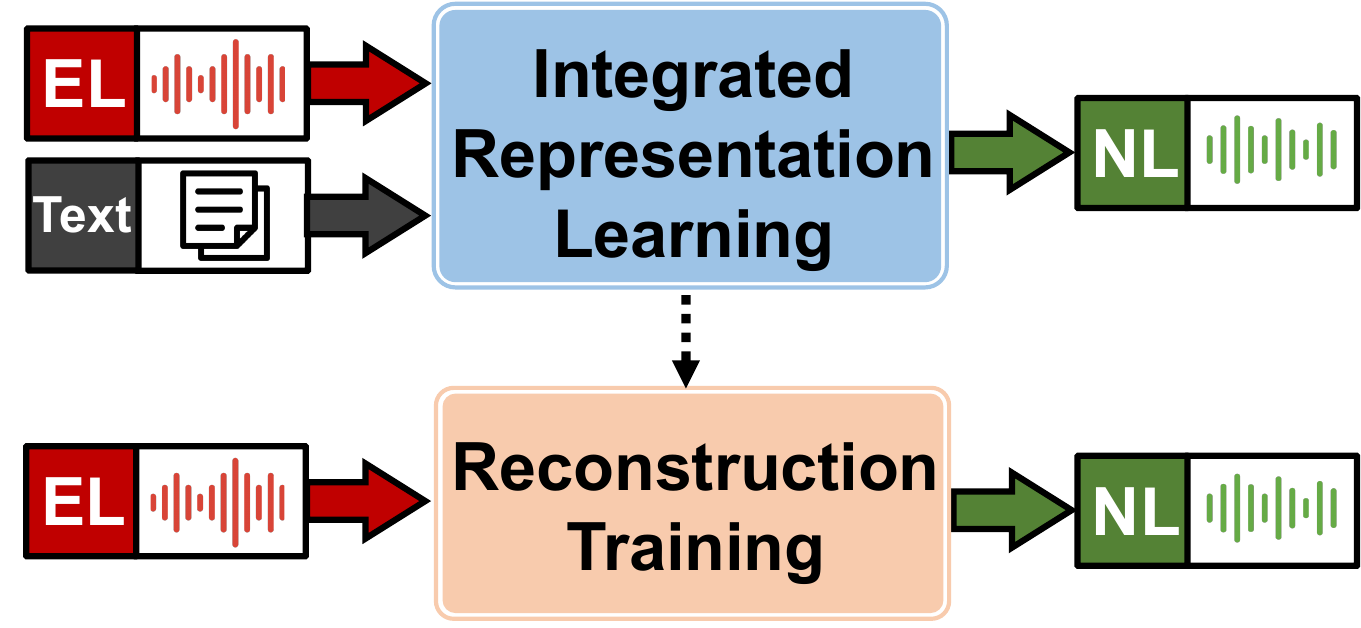}}
\caption{Basic concept of the proposed EL2SP framework with integrated speech–text representation learning and a reconstruction stage. “EL,” “NL,” and “Text” denote electrolaryngeal speech, normal speech, and the corresponding text representations, respectively.}
\label{fig3}
\end{figure}

On the basis of these observations, we propose a novel seq2seq-based speech--text representation learning framework for EL2SP. Unlike single speech-feature-based approaches, our approach integrates paired EL speech features and their corresponding text encodings within a seq2seq VC framework, which preserves a relatively simple architecture and inference scheme, and avoids reliance on large-scale multimodal training resources. With the stepping stone provided by our previous work~\cite{11254511}, we first construct a unified network consisting of a speech encoder, a text encoder, and a speech decoder. Through jointly learned speech--text representations, the network acquires more precise intermediate representations that better support mapping toward target normal speech. These representations are subsequently inherited to the final EL2SP system while preserving the original seq2seq VC architecture. The basic concept of the proposed framework is illustrated in Fig.~\ref{fig3}. More crucially, we extend the previous works by addressing two overlooked issues: 1) how to further enhance the integration of speech and text features, and 2) how to more appropriately transfer the integrated representations into the seq2seq EL2SP framework. We develop different systems and conduct a systematic comparative study. The main contributions are as follows:
\begin{itemize}
\setlength{\itemsep}{0pt}
\setlength{\parsep}{0pt}
\setlength{\parskip}{0pt}
\item For the first time, we present the EL2SP framework that jointly integrates speech and text features and provide an in-depth analysis of its effectiveness. We prepare essential components including the text encoder and speech decoder from a TTS model fine-tuned on target normal speech. This configuration produces linguistically enriched representations that facilitate smoother EL-to-normal speech mapping.

\item We propose three integration strategies based on text representations: 1) middle-level fusion, which combines speech and text encodings in intermediate space; 2) input-level fusion, which explicitly injects text encodings into speech encoder inputs; and 3) hybrid-level fusion, which unifies both mechanisms from levels 1) and 2). Each one outperforms seq2seq VC baselines~\cite{yen2021mandarin,11095635} and reveals a consistent performance hierarchy.

\item We introduce a straightforward yet effective autoencoder-style reconstruction to transfer integrated representations into the final EL2SP system. After removing text encoder, the speech encoder can approximate the encoding performance achieved speech--text inputs, enabling enhanced EL2SP within a single seq2seq VC framework. Furthermore, we refine the reconstruction objective, which incorporates an auxiliary loss based on integrated representations, into EL2SP training, yielding further improvements.

\item We develop different EL2SP systems using the proposed integration and reconstruction strategies and evaluate them on multiple small-scale EL2SP datasets. Experimental results show that, when augmented with TTS-generated synthetic data (SD) and corresponding text, our systems achieve the best performance.

\end{itemize}

\section{Proposed Methods}

\begin{figure*}[!t]
\centering{\includegraphics[width=1\linewidth]{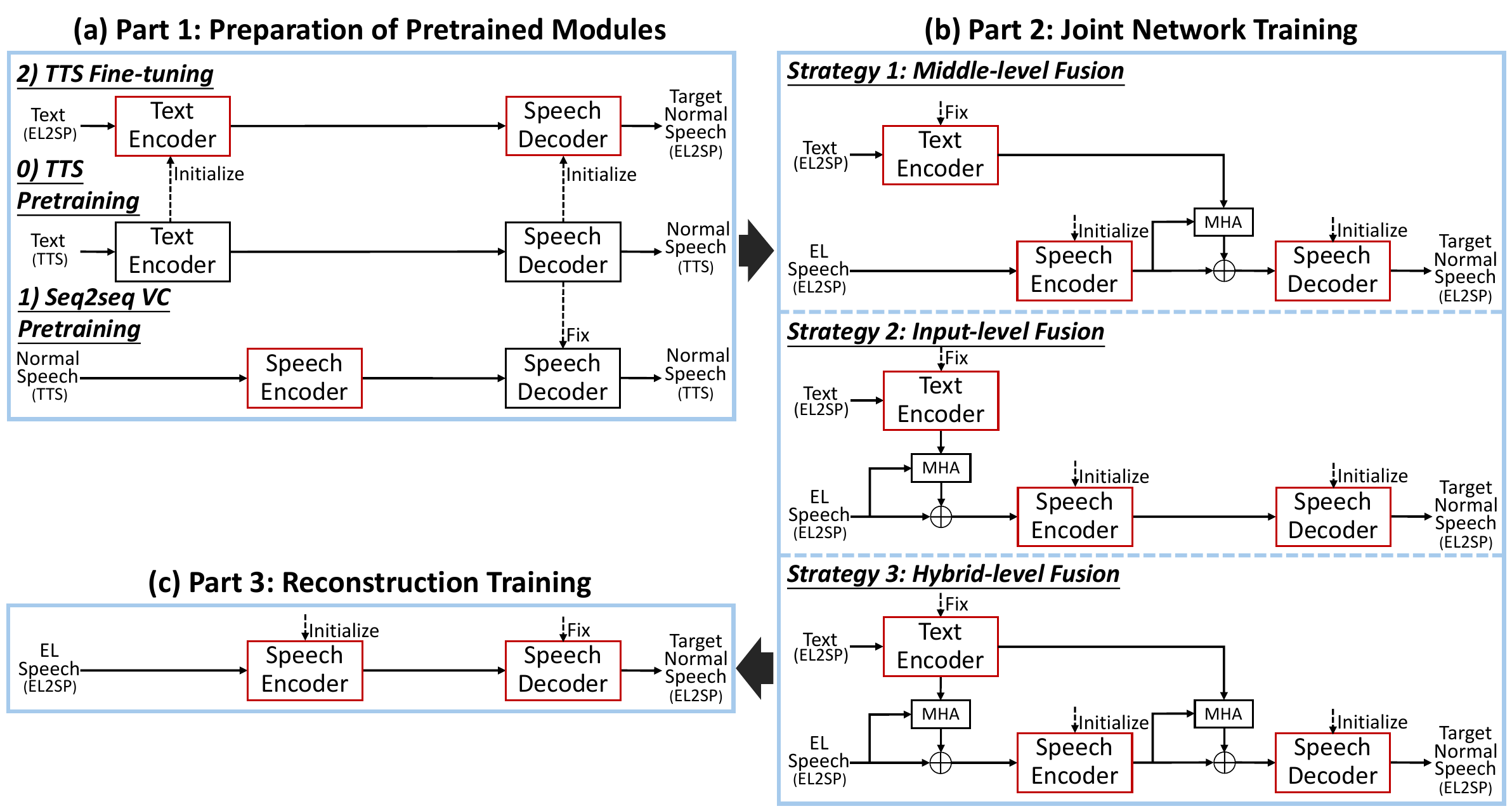}}
\caption{Overview of the training framework for the proposed representation learning methods. The pretrained modules from Part 1, namely the text encoder, speech encoder, and speech decoder (highlighted in red blocks), are inherited by the joint network in Part 2. Part 2 contains the middle-, input-, and hybrid-level fusion strategies, each serving as a different initialization for the reconstruction training in Part 3. MHA denotes the multi-head attention block used for cross attention.}
\label{fig4}
\end{figure*}

Recalling the basic concept described in the Introduction, the core idea of this study is to incorporate text representations into a seq2seq-based EL2SP framework to enhance modeling performance, while keeping the finalized EL2SP system practical. Specifically, the final system is expected to preserve the single original seq2seq VC architecture, without relying on any computationally expensive pretrained models or requiring additional text input during inference. To realize this, we propose a training framework for representation learning consisting of three parts, as illustrated in Fig.~\ref{fig4}. Part 1 in Section II-A introduces the pretraining strategies that use a TTS database together with a small-volume EL2SP dataset, thus providing the pretrained modules needed to initialize the subsequent joint network. Part 2 in Section II-B performs joint network training on the EL2SP dataset, during which different representation-fusion strategies are proposed to effectively learn intermediate speech--text representations. Part 3 in Section II-C conducts reconstruction training using only the speech data from the EL2SP dataset to derive the final EL2SP model. In this stage, the reconstruction objective is further refined to improve transfer of the integrated representations into the seq2seq framework. Since this study involved human subjects, the collection and use of the data were approved by the Ethics Committee of Nagoya University Hospital (Approval No.~2021-0232), and informed consent was obtained from all participants.

\subsection{Part 1: Preparation of Pretrained Modules}

As illustrated in Fig.~\ref{fig4}(a), Part 1 consists of two training branches that originate from the same Transformer-based~\cite{vaswani2017attention} TTS pretraining. These branches are designed to construct: 1) a pretrained seq2seq VC model, and 2) a phoneme-level TTS model for the target normal speech in EL2SP. In TTS pretraining stage, a large-scale normal-speaker TTS database is leveraged, allowing the encoder is well trained to extract linguistically pure representations, while the decoder learns an effective mapping from these representations to target speech features. 
\subsubsection{Seq2seq VC Pretraining} 
Based on our previous study~\cite{11095635}, we construct a one-to-one seq2seq VC model by employing Voice Transformer Network~\cite{huang2019voice} initialized through TTS pretraining. Since VC and TTS share the same decoding mechanism, the representation knowledge learned in TTS can be transferred to VC. In VC, the encoder must suppress speaker-dependent information in source speech to produce linguistically oriented representations. By contrast, TTS can more easily enforce such purely linguistic representations by using a sufficiently large corpus from an arbitrary single speaker with corresponding transcriptions. This reduces the data requirements and enables potential knowledge transfer from TTS to VC. During seq2seq VC pretraining, we remove the pretrained text encoder and introduce a new speech encoder. The pretrained speech decoder is kept frozen, the speech encoder is hence trained by minimizing a reconstruction loss using the same TTS corpus as both input and output. Because the frozen decoder has already adapted to a stable linguistic representation space, this procedure encourages the speech encoder to learn to extract similar linguistic representations directly from speech rather than from text.

\subsubsection{Target Normal TTS Fine-tuning}
The pretrained TTS model is fine-tuned using the target normal speech data and their corresponding transcripts from the EL2SP dataset. Although the limited fine-tuning data inevitably constrains the overall TTS performance, it still preserves the ability to map text representations to the target normal speech features, which is expected to facilitate smoother mapping process in the subsequent speech--text integration learning.

\subsection{Part 2: Speech--text-based Integrated Representation Learning}

Building upon the pretraining from Part 1, Part 2 constructs the joint network initialized with the speech encoder of the pretrained seq2seq VC model and the text encoder and speech decoder of the target normal TTS model. The objective of Part 2 is to obtain richer and more precise intermediate representations through jointly learning speech--text representations, thereby enabling a more effective mapping to the target normal speech. Here, EL speech data and their corresponding text data serve as dual inputs, while the target normal speech data are used as the single output. Since the text encoder receives the same type of text input as in Part 1, it is kept frozen during training to maintain consistency. Consequently, parameter updates in the joint network primarily focus on adapting the speech encoder to the newly introduced EL speech and the decoder to the fused representations. 

Let \(T=(t_1,t_2,...,t_l)\) represent a text feature sequence, where \(l\) denotes its length. We use the text encoder to obtain the text representations \(H_T\), which can be  formulated as
\begin{align}
  {H_T = \mathrm{TextEnc}(T)} \in \mathbb{R}^{l \times d},
\end{align}
where \(d\) represents the dimension of representations. Meanwhile, the input speech sequence is defined as \(S=(s_1,s_2,...,s_n)\) with \(n\) frames. To comprehensively investigate how speech and text representations can be effectively combined, we introduce three fusion strategies operating at different levels, as shown in Fig~\ref{fig4}(b), which are described in detail next. 

\subsubsection{Middle-level Fusion} 

This strategy combines speech and text encodings at an intermediate semantic level, providing a stable basis for semantic alignment.
The speech sequence is first processed by the speech encoder to obtain high-dimension speech representations \({H_S^{(\mathrm{MF})}}\), i.e.,
\begin{align}
  {{H_S^{(\mathrm{MF})}} = \mathrm{SpeechEnc}(S)} \in \mathbb{R}^{n \times d}.
\end{align}
Then, we combine \({H_S^{(\mathrm{MF})}}\) and \(H_T\) using a multi-head attention (MHA) mechanism in a cross-attention manner, where \({H_S^{(\mathrm{MF})}}\) acts as the query, and \(H_T\) serves as the key and value. Subsequently, a residual connection is applied to reintegrate \({H_S^{(\mathrm{MF})}}\). The corresponding integrated representations, \({H_\text{Fused}^{(\mathrm{MF})}}\), is formulated as
\begin{align}
  {{H_\text{Fused}^{(\mathrm{MF})}} = \mathrm{MHA}({H_S^{(\mathrm{MF})}}, H_T, H_T)+{H_S^{(\mathrm{MF})}}} \in \mathbb{R}^{n \times d}.
\end{align}
The MHA is defined as
\setlength{\jot}{3pt}
\begin{gather}
  {\mathrm{MHA}(Q,K,V)}=[\mathrm{head}_1,...,\mathrm{head}_h]W^{(O)}, \\
  {\mathrm{head}_i}=\mathrm{Att}(Q{W_i}^{(Q)},K{W_i}^{(K)},V{W_i}^{(V)}), \\
  {\mathrm{Att}(Q,K,V)=\mathrm{softmax} \left(\frac{QK^{\top}}{\sqrt{d_\mathrm{att}}}\right)V},
\end{gather}
where \(d_\mathrm{att}\) denotes the attention dimension, \(Q\), \(K\), and \(V\) are referred to as the query, key, and value, respectively. MHA uses \(h\) learned projections \({W}^{(Q)}\), \({W}^{(K)}\), \({W}^{(V)}\) to map the inputs to different heads, and then applies the \(\mathrm{Att}\) operation in parallel across the heads. The outputs from all heads are concatenated and projected with \({W}^{(O)}\).

\subsubsection{Input-level Fusion} 
This strategy performs early fusion by injecting the text encodings directly into the speech sequence before the speech encoder. Unlike middle-level fusion, this strategy allows the speech encoder to work with text-enriched representations from the beginning of its processing, which enables it to more explicitly internalize text-aligned cues. As a result, the encoder is expected to retain part of these text-informed patterns, providing a beneficial inductive bias for the final EL2SP model during the subsequent reconstruction training stage, where only speech inputs are used. Specifically, the speech sequence \(S\) and the text encoder outputs \(H_T\) are combined using MHA mechanism, where \(S\) serves as query, and \(H_T\) serves as the key and value. The attended features are added back to the original speech sequence through a residual connection. Then, the fused sequence is processed by the speech encoder for further representation learning. This process is formulated as
\begin{align}
H_S^{(\mathrm{IF})} = \mathrm{SpeechEnc}(\mathrm{MHA}(S,H_T,H_T)+S) \in \mathbb{R}^{n\times d}. 
\end{align}
Hence, the corresponding integrated representations, \({H_\text{Fused}^{(\mathrm{IF})}}\), can be directly represented as

\begin{align}
H_\text{Fused}^{(\mathrm{IF})} = H_S^{(\mathrm{IF})}.
\end{align}

\subsubsection{Hybrid-level Fusion} 

This strategy combines both middle- and input-level fusions to leverage their complementary strengths, where the text encoder outputs \(H_T\) are fed at both the input- and middle-level fusion positions. Following the input-level fusion, the speech sequence \(S\) first attends to \(H_T\) using MHA mechanism, and the outputs of MHA are added back via a residual connection. The fused sequence is then fed into the speech encoder to obtain high-dimension speech representations \({H_S^{(\mathrm{HF})}}\), i.e., 
\begin{align}
  {H_S^{(\mathrm{HF})}} = {H_S^{(\mathrm{IF})}}.
\end{align}

Subsequently, similar to the middle-level fusion, a second MHA module is applied to integrate \({H_S^{(\mathrm{HF})}}\) with \(H_T\) at the semantic level, again using a residual connection to stabilize the integration. Therefore, the finalized integrated representations \({H_\text{Fused}^{(\mathrm{HF})}}\) are formulated as
\begin{align}
  {{H_\text{Fused}^{(\mathrm{HF})}} = \mathrm{MHA}({H_S^{(\mathrm{HF})}}, H_T, H_T)+{H_S^{(\mathrm{HF})}}} \in \mathbb{R}^{n \times d}.
\end{align}

In summary, the three fusion strategies offer different ways to build integrated representations by incorporating text information while retaining the complete speech representations. The updated speech decoder is expected to adapt to these integrated representations and thus can decode them into target speech.

\subsection{Part 3: Reconstruction Training} 

Based on the proposed fusion strategies in Part 2, Part 3 aims to inherit these integrated representations while preserving the original seq2seq VC architecture to ensure compatibility with the EL2SP scenario. As illustrated in Fig~\ref{fig4}(c), we remove the text encoder and initialize both the speech encoder and decoder with the parameters learned in Part 2. An autoencoder-style reconstruction training is then performed, in which the decoder is fixed and only the speech encoder is updated using the same EL speech \(S\) as input. This procedure finalizes the EL2SP model without relying on text information.

As described previously, the decoder has adapted to the integrated representations during Part 2. Let \(c \in \{\mathrm{MF}, \mathrm{IF}, \mathrm{HF}\}\) denote the fusion strategy. In Part 3, the speech encoder is updated to reconstruct representations \(\tilde{H}_S^{(c)}\) that closely approximate the integrated targets \(H_\text{Fused}^{(c)}\), thereby maintaining consistency with the decoder’s established mapping behavior while only consuming \(S\). This process is formulated as
\begin{equation}
\begin{split}
  \tilde{H}_S^{(c)} = \mathrm{SpeechEnc}(S) \in \mathbb{R}^{n \times d}
  \approx H_\text{Fused}^{(c)} \in \mathbb{R}^{n \times d}, \\
  \text{where} \quad c \in \{\mathrm{MF}, \mathrm{IF}, \mathrm{HF}\}.
\end{split}
\end{equation}
Thus, we can transfer more effective integrated representations to the EL2SP framework, enhancing the overall performance. 

Here, we adopt several standard objectives used in seq2seq VC: an L1 loss on the final acoustic outputs, a weighted binary cross-entropy loss for the stop-token prediction, and a guided attention loss applied for diagonal alignment \cite{tachibana2018efficiently,huang2019voice} in Transformer. These correspond to \(\mathcal{L}_{\text{seq}}\), \(\mathcal{L}_{\text{token}}\), \(\mathcal{L}_{\text{ga}}\) respectively. The total loss \(\mathcal{L}_{\text{total}}\) is formulated as
\begin{gather}
  \mathcal{L}_{\text{total}} = \mathcal{L}_{\text{seq}}+\mathcal{L}_{\text{token}}+\mathcal{L}_{\text{ga}}.
\end{gather}

Furthermore, to better exploit the effectiveness of integrated representations, we extend the typical training objective \(\mathcal{L}_{\text{total}}\) by an integrated-representation–guided reconstruction loss, denoted \(\mathcal{L}_{\text{rec}}^{(c)}\). This loss is formulated as an L1 distance between the reconstructed representations \(\tilde{H}_S^{(c)}\) and the integrated targets \(H_\text{Fused}^{(c)}\). In this circumstance, the extended loss \(\mathcal{L}_{\text{total-rec}}^{(c)}\) is expressed as
\begin{gather}
  \mathcal{L}_{\text{rec}}^{(c)} = \left\lVert \tilde{H}_S^{(c)} - H_\text{Fused}^{(c)} \right\rVert_{1}, \\
  \mathcal{L}_{\text{total-rec}}^{(c)} = \mathcal{L}_{\text{seq}}+\mathcal{L}_{\text{token}}+\mathcal{L}_{\text{ga}}+\lambda \mathcal{L}_{\text{rec}}^{(c)},
\end{gather}
where \(\lambda\) is a predefined coefficient that controls the strength of the auxiliary alignment. With this formulation, the reconstruction training is augmented by encouraging the model to attend to \(H_\text{Fused}^{(c)}\) from Part 2, encouraging more explicit alignment learning for the speech encoder. Note that \(H_\text{Fused}^{(c)}\) is only used for loss formulation.

\section{Experimental Protocols}

\subsection{System Building}

By combining different fusion strategies and training objectives from the proposed framework, we construct a series of EL2SP systems. To validate their effectiveness, we compare them against two baseline systems, both of which operate solely on speech-based representations. For a fair comparison, all the systems use the same seq2seq VC pretraining setup. A summary that distinguishes these systems is provided in Table~\ref{tab:systems_all}.

\begin{table*}[t]\LARGE
\centering
\caption{Comparable systems under speech- and integrated-representation-based categories. Aside from the baselines, nine systems are constructed based on three fusion strategies on our proposed methods: middle-level fusion (MF), input-level fusion (IF), and hybrid-level fusion (HF). For each strategy, three variants (``-1'', ``-2'', ``-3'') are designed, differing in whether synthetic data (SD) and the extended loss function are incorporated. Accordingly, the proposed systems are denoted as P-MF-1/2/3, P-IF-1/2/3, and P-HF-1/2/3, respectively.}

\label{tab:systems_all}
\resizebox{\textwidth}{!}{%
\begin{tabular}{c|cc|ccccccccc}
\toprule[1.5pt]
\multicolumn{12}{c}{\textbf{Comparative categories}} \\
\midrule
 & \multicolumn{2}{c|}{\textbf{Speech-representation-based}}
 & \multicolumn{9}{c}{\textbf{Integrated-representation-based}} \\
\midrule[1pt]
\textbf{System building}
 & \multicolumn{1}{c|}{Baseline 1~\cite{yen2021mandarin}}
 & Baseline 2~\cite{11095635}
 & \multicolumn{1}{c|}{P-MF-1}
 & \multicolumn{1}{c|}{P-MF-2}
 & \multicolumn{1}{c|}{P-MF-3}
 & \multicolumn{1}{c|}{P-IF-1}
 & \multicolumn{1}{c|}{P-IF-2}
 & \multicolumn{1}{c|}{P-IF-3}
 & \multicolumn{1}{c|}{P-HF-1}
 & \multicolumn{1}{c|}{P-HF-2}
 & P-HF-3 \\
\midrule[1.5pt]
\textbf{\makecell[c]{Training\\framework}}
 & \multicolumn{1}{c|}{\makecell{Pretraining\\\&\\fine-tuning}}
 & \makecell{Pretraining\\\& two-stage\\fine-tuning}
 & \multicolumn{3}{c|}{\makecell{Parts 1 to 3\\(Middle-level fusion)}}
 & \multicolumn{3}{c|}{\makecell{Parts 1 to 3\\(Input-level fusion)}}
 & \multicolumn{3}{c}{\makecell{Parts 1 to 3\\(Hybrid-level fusion)}} \\
\midrule
\textbf{Use synthetic data?}
 & \multicolumn{1}{c|}{No}
 & Yes
 & \multicolumn{1}{c|}{No}
 & \multicolumn{1}{c|}{Yes}
 & \multicolumn{1}{c|}{Yes}
 & \multicolumn{1}{c|}{No}
 & \multicolumn{1}{c|}{Yes}
 & \multicolumn{1}{c|}{Yes}
 & \multicolumn{1}{c|}{No}
 & \multicolumn{1}{c|}{Yes}
 & Yes \\
\midrule
\textbf{Use text data?}
 & \multicolumn{1}{c|}{No}
 & No
 & \multicolumn{1}{c|}{Yes}
 & \multicolumn{1}{c|}{Yes}
 & \multicolumn{1}{c|}{Yes}
 & \multicolumn{1}{c|}{Yes}
 & \multicolumn{1}{c|}{Yes}
 & \multicolumn{1}{c|}{Yes}
 & \multicolumn{1}{c|}{Yes}
 & \multicolumn{1}{c|}{Yes}
 & Yes \\
\midrule
\textbf{Use $\mathcal{L}_{\text{rec}}^{(c)}$?}  
 & \multicolumn{1}{c|}{---}
 & ---
 & \multicolumn{1}{c|}{No}
 & \multicolumn{1}{c|}{No}
 & \multicolumn{1}{c|}{Yes}
 & \multicolumn{1}{c|}{No}
 & \multicolumn{1}{c|}{No}
 & \multicolumn{1}{c|}{Yes}
 & \multicolumn{1}{c|}{No}
 & \multicolumn{1}{c|}{No}
 & Yes \\
\bottomrule[1.5pt]
\end{tabular}%
}
\end{table*}

\begin{itemize}
    \item \textbf{Baseline 1:} It follows a conventional pretraining–fine-tuning pipeline similar to that in~\cite{yen2021mandarin}. The pretrained seq2seq VC model is directly fine-tuned on the original EL2SP dataset without any additional data augmentation.

    \item \textbf{Baseline 2:} Similar to the method in \cite{11095635}, it incorporates a two-stage fine-tuning scheme with parallel synthetic data (SD). To obtain parallel SD, we fine-tune the same pretrained TTS model used in Part 1 on the EL speech data to obtain an EL TTS model. By feeding an external text dataset into both the target-normal and EL TTS models, a larger-scale set of parallel SD is generated. The original EL2SP data and parallel SD are then jointly utilized in the first-stage fine-tuning. Since the quality of SD is imperfect, the second-stage fine-tuning is performed exclusively on the original EL2SP dataset to refine the model parameters and achieve further improvements.

    \item \textbf{P-MF-1, P-IF-1, P-HF-1 (``-1'' systems):} They are constructed by fully implementing Parts 1 to 3 using only the original EL2SP dataset (and its paired text for Part 2), while adopting the standard training objective as in Equation (12). Their differences arise from the fusion strategy employed in Part 2: middle-level fusion for P-MF-1, input-level fusion for P-IF-1, and hybrid-level fusion for P-HF-1.
    
    \item \textbf{P-MF-2, P-IF-2, P-HF-2 (``-2'' systems):} These systems extend their respective ``-1'' systems by incorporating SD during training, inspired by the data augmentation used in Baseline 2. In Part 2, the joint network is trained using both the original EL2SP dataset and the parallel SD, together with their corresponding text. In Part 3, only the original EL2SP data are used for reconstruction training to avoid potential degradation caused by imperfections in SD. Other than the use of SD, these systems follow the same fusion-based distinctions and the same training objective as their “-1” counterparts.
    
    \item \textbf{P-MF-3, P-IF-3, and P-HF-3 (``-3'' systems):} These systems build upon their corresponding “-2” systems, retaining settings consistent with their use of SD and fusion-based distinctions, while extending the training objective defined in Equation (14), by incorporating the integrated-representation-guided loss term \(\mathcal{L}_\text{rec}^{(c)}\) as an auxiliary supervision signal in Part 3. This enables the speech encoder to better align with the integrated representations learned in Part 2. Overall, these systems represent a enhancement relative to their ``-2'' and ``-1'' counterparts.

\end{itemize}
\vspace{-3mm}
\subsection{Datasets}
The proposed systems rely on two types of datasets, which are introduced as follows.

\subsubsection{TTS Database}
To accomplish the TTS pretraining and seq2seq VC pretraining of Part 1 described in Section II-A, we utilized the single-speaker Japanese JSUT corpus~\cite{sonobe2017jsut}, which contains 7,296 utterances, approximately 10 hours long. In addition, the JSUT transcriptions were used to generate SD. Both the JSUT transcriptions and the generated SD were used as training inputs for the joint network in Part 2.

\subsubsection{Original EL2SP Datasets}
To comprehensively validate the proposed EL2SP systems, we built four different small-scale EL2SP datasets. This dataset design represents a practical low-resource EL2SP scenario, where collecting a large amount of speech data from patients is difficult in practice due to strict recording conditions and the fragile physical functions of patients~\cite{11095635,10533680}. Among these dataset, two of them involve EL speech recorded from laryngectomees, referred to as the Patient-1 and Patient-2 datasets, and the other two were constructed using simulated EL speech produced by healthy speakers, referred to as the Pseudo-Patient-1 and Pseudo-Patient-2 datasets. All recordings were collected under uniform conditions in a professional soundproof booth using a Shure SM58 microphone, and a Roland Rubix 22 interface with Audacity recording software. Each dataset was independently constructed by native Japanese speakers and contains its own set of utterances, providing diverse and complementary evaluation scenarios. 

\begin{itemize}
    \item \textbf{Patient-1 dataset:} This dataset was a semi-parallel corpus, containing 200 EL utterances under 10 minutes and 413 normal speech utterances around 20 minutes. The EL speech was produced by a laryngectomee using an electrolarynx. Since the patient underwent a total laryngectomy for cervical esophageal cancer and no pre-surgery normal speech could be obtained, a healthy male speaker recorded the normal utterances under the same recording condition for a normal speech reference. This dataset simulated a practical scenario where the patient’s healthy voice is no longer accessible.

    \item \textbf{Patient-2 dataset:} This dataset was also constructed under a semi-parallel setting, containing 643 EL utterances near 30 minutes and 373 normal speech utterances about 18 minutes. The EL recordings were collected from a laryngectomee who had been treated for advanced hypopharyngeal cancer. The treatment involved a total removal of larynx and subsequent reconstruction using a jejunal graft. The patient’s normal voice could still be recorded prior to surgery, as the tumor did not impair vocal-fold function at that stage. This dataset therefore represented a condition where preoperative normal speech from a laryngectomee is obtainable.

    \item \textbf{Patient-3 dataset:} This dataset was a parallel corpus containing 300 EL--normal utterance pairs totaling approximately 15 minutes. The EL and normal recordings were collected from a patient who underwent total pharyngolaryngectomy with free jejunal flap reconstruction for locally advanced hypopharyngeal cancer. Similar to the Patient-2 dataset, the vocal cords of this patient were not significantly affected before surgery, allowing the normal voice to be recorded preoperatively.
    
    \item \textbf{Pseudo-Patient-1/2 datasets:} Because collecting real-patient datasets is difficult, the pseudo-patient datasets aim to provide additional evaluation scenarios under limited-data conditions. These two datasets were constructed from healthy speakers who produced both simulated EL speech using an electrolarynx and their natural voices. In particular, before recording the simulated EL speech, the speakers carefully practiced producing EL speech in a patient-like manner. Therefore, these pseudo-patient datasets still retain essential EL speech characteristics, and thus allow us to further validate the robustness of the proposed systems under multiple small-scale EL2SP settings. Similar dataset setting can be viewed in the related works~\cite{kobayashi2018electrolaryngeal, 11095635}. The Pseudo-patient-1 dataset contained 413 parallel EL--normal utterance pairs totaling about 20 minutes, while the Pseudo-patient-2 dataset consisted of 200 pairs within 10 minutes.
\end{itemize}

These four EL2SP datasets adopted a consistent data split: 20 parallel utterances for the development set and 40 for the test set, while the remaining utterances were used for training. For Baseline 1, P-MF-1, P-IF-1, and P-HF-1, the Pseudo-patient-1 and -2 datasets used 353 and 140 parallel training utterances, respectively, while the Patient-1 and -2 datasets used only their remaining parallel portions, i.e., 140 and 313 training utterances, respectively. 

When implementing Baseline 2, P-MF-2/3, P-IF-2/3, and P-HF-2/3 that leveraged SD, the 7,296 external TTS-generated parallel SD was additionally used. It is worth noting that, to fully leverage all feasible original data in the Patient-1 and -2 datasets, we also used fine-tuned TTS models to add corresponding EL SD (213) for the Patient-1 dataset and normal SD (270) for the Patient-2 dataset, respectively. In their final training stages, these systems again used the original parallel data, with 140 and 313 pairs for the Patient-1 and -2 datasets, respectively.
\vspace{-3mm}
\subsection{Implementations}

The proposed systems were implemented using the ESPnet~\cite{watanabe2018espnet} toolkit. All speech signals were resampled to 24 kHz, and 80-dimensional mel-filterbank features were extracted using a 2048-point Fast Fourier Transform (FFT) and a 300-sample frame shift. The input text features for the TTS models consisted of Japanese phonemes and pause symbols. Both the Transformer-based TTS and seq2seq VC models followed similar configurations, employing six encoder and six decoder layers with four attention heads per layer. The hidden dimensions were set to 384 for attention and 1,536 for feed-forward layers. The decoder reduction factor was set to 3, and layer normalization was applied before the encoder blocks. In addition, in the joint network of the proposed methods, all MHA modules used in the fusion strategies also employed four attention heads.

The LAMB~\cite{you2019large} and Noam~\cite{vaswani2017attention} optimizers were used for VC- and TTS-related training, respectively, with an initial learning rate of 0.001 and a gradient accumulation step of 1. Across all proposed systems, the finalized EL2SP models inherited the same seq2seq VC framework and thus maintained a model size of 30.4 million trainable parameters, identical to that of the baseline systems. To reconstruct waveforms for SD and EL2SP outputs, eight speaker-dependent Parallel WaveGAN vocoders~\cite{yamamoto2020parallel} were trained across the four datasets based on the default configurations provided\footnote{\href{https://github.com/kan-bayashi/ParallelWaveGAN}{\scriptsize\url{https://github.com/kan-bayashi/ParallelWaveGAN}}}.

For systems incorporating the extended loss function, the weighting coefficient~\(\lambda\) was set to 0.01, ensuring the dominance of the original reconstruction objective while the auxiliary supervision acts as a light semantic constraint that stabilizes model training.

\subsection{Evaluation Metrics}
\subsubsection{Objective Evaluation Metrics}
We adopted four objective evaluation metrics to assess different aspects of the proposed methods.

\begin{itemize}
    \item Mel-Cepstrum Distortion \cite{kubichek1993mel} (MCD, in dB): This metric measures the spectral envelope distortion by computing the L2 distance between the mel-cepstral coefficients (MCCs) of the generated and ground-truth samples, after aligning the MCC sequences using dynamic time warping (DTW). It is defined as
    \begin{equation}
    \text{MCD} = \frac{10}{\ln 10} \sqrt{2 \sum_{i=1}^{D} \left( c_i^{\text{(ref)}} - c_i^{\text{(gen)}} \right)^2 },
    \end{equation}
    where $D$ denotes the feature dimension of the MCCs, and $c_i^{\text{(gen)}}$ and $c_i^{\text{(ref)}}$ represent the $i$-th dimensional coefficients of the generated and target original MCCs, respectively. A lower MCD indicates less spectral envelope distortion and therefore higher overall quality and fidelity of the converted speech.
    
    \item Character Error Rate (CER, in \%): This metric evaluates the intelligibility by comparing the recognized character sequence of the converted speech with the reference text. Given that the proposed methods focused on Japanese language, CER provides a character-level measurement of linguistic consistency. To calculate CER, we employed a pretrained ASR model provided in~\cite{violeta2022investigating}. A lower CER value represents higher intelligibility accuracy.
    
    \item Log F0 Root Mean Square Error (F0 RMSE): This metric is a frame-level objective measure for evaluating the accuracy of pitch reconstruction. Following the same DTW-based time alignment as in the MCD calculation, it computes the RMSE between the logarithmic F0 values of the generated and ground-truth speech. It is defined as:
        \begin{equation}
        \text{F0 RMSE} = \sqrt{\frac{1}{N} \sum_{i=1}^{N} \left( \log f_0^{\text{(gen)}}(i) - \log f_0^{\text{(ref)}}(i) \right)^2 },
        \end{equation}
    where $N$ is the total number of voiced frames, and $f_0^{\text{(gen)}}(i)$ and $f_0^{\text{(ref)}}(i)$ are the F0 values of the generated and reference speech at frame $i$, respectively. A lower F0 RMSE value indicates more precise pitch estimation.

    \item Log F0 Correlation (F0 CORR): This metric evaluates the similarity of the overall log-F0 contours between the converted and ground-truth speech with the Pearson correlation coefficient, after applying the same DTW-based time alignment as in the MCD calculation. A higher value (closer to 1) indicates a closer pitch alignment.
    
\end{itemize}
\subsubsection{Subjective Evaluation Metrics}

To assess perceptual performance, we conducted Mean Opinion Score (MOS)~\cite{international1996methods} evaluations from two perspectives: speech naturalness and speech intelligibility.

\begin{itemize}
\item Naturalness Evaluation: Participants were asked to rate how natural each generated sample sounded. The scoring criteria were given on a five-point scale from one (completely unnatural) to five (completely natural). On the evaluation interface, each sample was presented with the following selectable options: (1) \textit{Bad}, (2) \textit{Poor}, (3) \textit{Fair}, (4) \textit{Good}, and (5) \textit{Excellent}, where a higher score indicates greater perceptual naturalness.

\item Intelligibility Evaluation: With evaluation settings similar to those in~\cite{nishio2026voice,zhou2026cyclegan}, listeners evaluated how clearly each speech sample conveyed the intended message. Unlike the naturalness test, participants were provided with both the speech sample and the corresponding text as reference. Intelligibility was rated on a five-point MOS scale, where 5 denotes \textit{Very intelligible}, 4 \textit{Mostly intelligible}, 3 \textit{Partially intelligible}, 2 \textit{Not very intelligible}, and 1 \textit{Not intelligible at all}.
\end{itemize}

For both evaluations, fifteen samples were randomly selected from each system. To provide clear perceptual anchors, the original EL speech and the target normal recordings were also included, serving as lower and upper bounds. The subjective tests followed standard protocols in VC and EL2SP research by collecting perceptual judgments from general listeners rather than from a recruited expert panel~\cite{kobayashi2018electrolaryngeal,11095635,nishio2026voice}. Twenty general adult native Japanese listeners were recruited. This setting is aligned with the intended application of the proposed system, as the goal of EL2SP is to improve communication in broad real-world scenarios involving the general public. Audio samples are available online\footnote{\href{https://silenticymoon.github.io/TBMEdemo/}{\scriptsize\url{https://silenticymoon.github.io/TBMEdemo/}}}.

\section{Experimental Results}

\subsection{Objective Evaluation Results}
The comparative results of Baselines 1 and 2, as well as all the proposed systems, including ``-1'' systems (P-MF-1, P-IF-1, P-HF-1), ``-2'' systems (P-MF-2, P-IF-2, P-HF-2), and ``-3'' systems (P-MF-3, P-IF-3, P-HF-3), across the five datasets are summarized in Tables~\ref{tab:table1},~\ref{tab:table2},~\ref{tab:table-p3},~\ref{tab:table3}, and~\ref{tab:table4}.

\newcommand{\symdag}{$^{\dagger}$}
\newcommand{\symnone}{$^{\phantom{\dagger}}$}
\begin{table}[h]\large
\centering
\caption{Objective evaluation results based on Patient-1 dataset.}
\label{tab:table1}
\resizebox{\linewidth}{!}{
\begin{tabular}{c|c|c|c|c}
\toprule[1.5pt]
\multirow{2}{*}{\textbf{Systems}} & \multicolumn{4}{c}{\textbf{Objective Evaluation Metrics}} \\ 
\cmidrule(lr){2-5}
 & \textbf{MCD} ($\downarrow$) & \textbf{CER} ($\downarrow$) & \textbf{F0 RMSE} ($\downarrow$) & \textbf{F0 CORR} ($\uparrow$) \\ 
\midrule
\textbf{Baseline 1} & 7.17\symdag & 41.3\symdag & 0.26\symdag & 0.65\symdag \\[2.5pt] 
\textbf{Baseline 2} & 6.24\symdag & 23.3\symdag & 0.25\symdag & 0.67\symdag \\[2.5pt]
\addlinespace[1pt] \cline{1-5} \addlinespace[2.5pt]
\textbf{P-MF-1} & 6.21\symdag & 31.5\symdag & 0.24\symdag & 0.67\symdag \\[2.5pt] 
\textbf{P-MF-2} & 6.04\symdag & 22.1\symdag & 0.23\symnone & 0.69\symdag \\[2.5pt]
\textbf{P-MF-3} & 6.00\symdag & 20.7\symdag & \textbf{0.22}\symnone & 0.69\symdag \\[2.5pt]
\addlinespace[1pt] \cline{1-5} \addlinespace[2.5pt]
\textbf{P-IF-1} & 6.16\symdag & 28.4\symdag & 0.24\symdag & 0.70\symdag \\[2.5pt]
\textbf{P-IF-2} & 5.89\symdag & 19.7\symnone & 0.23\symdag & 0.69\symdag \\[2.5pt]
\textbf{P-IF-3} & 5.87\symdag & 19.1\symnone & \textbf{0.22}\symnone & 0.67\symdag \\[2.5pt]
\addlinespace[1pt] \cline{1-5} \addlinespace[2.5pt]
\textbf{P-HF-1} & 6.11\symdag & 27.9\symdag & \textbf{0.22}\symnone & 0.66\symdag \\[2.5pt]
\textbf{P-HF-2} & 5.75\symnone & 20.0\symdag & 0.23\symnone & 0.70\symdag \\[2.5pt]
\textbf{P-HF-3} & \textbf{5.74}\symnone & \textbf{18.4}\symnone & \textbf{0.22}\symnone & \textbf{0.73}\symnone \\[2.5pt]
\bottomrule[1.5pt]
\end{tabular}}
\vspace{2pt}
\begin{flushleft}
\footnotesize $^{\dagger}$ indicates a statistically significant difference compared with P-HF-3 (\( \text{p} < 0.05 \)).
\end{flushleft}
\end{table}

\begin{table}[!t]\large
\centering
\caption{Objective evaluation results based on Patient-2 dataset.}
\label{tab:table2}
\resizebox{\linewidth}{!}{
\begin{tabular}{c|c|c|c|c}
\toprule[1.5pt]
\multirow{2}{*}{\textbf{Systems}} & \multicolumn{4}{c}{\textbf{Objective Evaluation Metrics}} \\ 
\cmidrule(lr){2-5}
 & \textbf{MCD} ($\downarrow$) & \textbf{CER} ($\downarrow$) & \textbf{F0 RMSE} ($\downarrow$) & \textbf{F0 CORR} ($\uparrow$) \\ 
\midrule
\textbf{Baseline 1} & 8.27\symdag & 45.7\symdag & 0.46\symdag & 0.84\symdag \\[2.5pt]
\textbf{Baseline 2} & 6.38\symdag & 26.2\symdag & 0.41\symdag & 0.87\symnone \\[2.5pt]
\addlinespace[1pt] \cline{1-5} \addlinespace[2.5pt]
\textbf{P-MF-1} & 7.04\symdag & 35.8\symdag & 0.44\symdag & 0.86\symnone \\[2.5pt]
\textbf{P-MF-2} & 6.23\symdag & 24.5\symdag & 0.40\symdag & 0.86\symnone \\[2.5pt]
\textbf{P-MF-3} & 6.21\symdag & 23.1\symdag & 0.41\symdag & 0.86\symnone \\[2.5pt]
\addlinespace[1pt] \cline{1-5} \addlinespace[2.5pt]
\textbf{P-IF-1} & 6.96\symdag & 34.8\symdag & 0.45\symdag & 0.85\symdag \\[2.5pt]
\textbf{P-IF-2} & 6.18\symdag & 21.2\symdag & 0.39\symnone & 0.87\symnone \\[2.5pt]
\textbf{P-IF-3} & 6.17\symdag & 20.6\symnone & 0.39\symnone & 0.86\symnone \\[2.5pt]
\addlinespace[1pt] \cline{1-5} \addlinespace[2.5pt]
\textbf{P-HF-1} & 6.95\symdag & 34.1\symdag & 0.46\symdag & 0.86\symdag \\[2.5pt]
\textbf{P-HF-2} & 6.13\symdag & 20.7\symnone & 0.39\symnone & \textbf{0.88}\symnone \\[2.5pt]
\textbf{P-HF-3} & \textbf{6.09}\symnone & \textbf{19.7}\symnone & \textbf{0.38}\symnone & 0.87\symnone \\[2.5pt]
\bottomrule[1.5pt]
\end{tabular}}
\vspace{2pt}
\begin{flushleft}
\footnotesize $^{\dagger}$ indicates a statistically significant difference compared with P-HF-3 (\( \text{p} < 0.05 \)).
\end{flushleft}
\end{table}

\begin{table}[!t]\large
\centering
\caption{Objective evaluation results based on Patient-3 dataset.}
\label{tab:table-p3}
\resizebox{\linewidth}{!}{
\begin{tabular}{c|c|c|c|c}
\toprule[1.5pt]
\multirow{2}{*}{\textbf{Systems}} & \multicolumn{4}{c}{\textbf{Objective Evaluation Metrics}} \\ 
\cmidrule(lr){2-5}
 & \textbf{MCD} ($\downarrow$) & \textbf{CER} ($\downarrow$) & \textbf{F0 RMSE} ($\downarrow$) & \textbf{F0 CORR} ($\uparrow$) \\ 
\midrule
\textbf{Baseline 1} & 7.96\symdag & 49.1\symdag & 0.35\symdag & 0.87\symdag \\[2.5pt]
\textbf{Baseline 2} & 7.42\symdag & 37.0\symdag & 0.35\symdag & 0.89\symdag \\[2.5pt]
\addlinespace[1pt] \cline{1-5} \addlinespace[2.5pt]
\textbf{P-MF-1} & 7.64\symdag & 46.5\symdag & 0.34\symdag & 0.90\symdag \\[2.5pt]
\textbf{P-MF-2} & 7.35\symdag & 35.5\symdag & 0.33\symnone & 0.89\symdag \\[2.5pt]
\textbf{P-MF-3} & 7.32\symdag & 33.5\symdag & 0.34\symdag & 0.89\symdag \\[2.5pt]
\addlinespace[1pt] \cline{1-5} \addlinespace[2.5pt]
\textbf{P-IF-1} & 7.59\symdag & 43.3\symdag & 0.33\symnone & 0.91\symnone \\[2.5pt]
\textbf{P-IF-2} & 7.30\symdag & 31.5\symdag & \textbf{0.32}\symnone & 0.90\symdag \\[2.5pt]
\textbf{P-IF-3} & 7.25\symdag & 30.2\symdag & 0.34\symdag & 0.91\symdag \\[2.5pt]
\addlinespace[1pt] \cline{1-5} \addlinespace[2.5pt]
\textbf{P-HF-1} & 7.54\symdag & 40.8\symdag & \textbf{0.32}\symnone & 0.90\symdag \\[2.5pt]
\textbf{P-HF-2} & 7.21\symnone & 29.0\symdag & 0.34\symdag & 0.90\symdag \\[2.5pt]
\textbf{P-HF-3} & \textbf{7.20}\symnone & \textbf{27.0}\symnone & \textbf{0.32}\symnone & \textbf{0.92}\symnone \\[2.5pt]
\bottomrule[1.5pt]
\end{tabular}}
\vspace{2pt}
\begin{flushleft}
\footnotesize $^{\dagger}$ indicates a statistically significant difference compared with P-HF-3 (\( \text{p} < 0.05 \)).
\end{flushleft}
\end{table}

\begin{table}[!t]\large
\centering
\caption{Objective evaluation results based on Pseudo-patient-1 dataset.}
\label{tab:table3}
\resizebox{\linewidth}{!}{
\begin{tabular}{c|c|c|c|c}
\toprule[1.5pt]
\multirow{2}{*}{\textbf{Systems}} & \multicolumn{4}{c}{\textbf{Objective Evaluation Metrics}} \\ 
\cmidrule(lr){2-5}
 & \textbf{MCD} ($\downarrow$) & \textbf{CER} ($\downarrow$) & \textbf{F0 RMSE} ($\downarrow$) & \textbf{F0 CORR} ($\uparrow$) \\ 
\midrule
\textbf{Baseline 1} & 6.37\symdag & 51.4\symdag & 0.27\symdag & 0.68\symdag \\[2.5pt]
\textbf{Baseline 2} & 5.77\symdag & 34.7\symdag & 0.24\symdag & 0.70\symdag \\[2.5pt]
\addlinespace[1pt] \cline{1-5} \addlinespace[2.5pt]
\textbf{P-MF-1} & 6.04\symdag & 48.7\symdag & 0.23\symdag & \textbf{0.73}\symnone \\[2.5pt]
\textbf{P-MF-2} & 5.64\symdag & 34.2\symdag & 0.24\symdag & 0.72\symnone \\[2.5pt]
\textbf{P-MF-3} & 5.63\symdag & 33.1\symdag & 0.23\symnone & 0.72\symnone \\[2.5pt]
\addlinespace[1pt] \cline{1-5} \addlinespace[2.5pt]
\textbf{P-IF-1} & 6.03\symdag & 46.2\symdag & 0.24\symdag & \textbf{0.73}\symnone \\[2.5pt]
\textbf{P-IF-2} & 5.55\symdag & 33.5\symdag & 0.23\symdag & 0.72\symnone \\[2.5pt]
\textbf{P-IF-3} & 5.51\symdag & 33.0\symdag & 0.23\symnone & 0.72\symnone \\[2.5pt]
\addlinespace[1pt] \cline{1-5} \addlinespace[2.5pt]
\textbf{P-HF-1} & 5.95\symdag & 45.6\symdag & 0.23\symnone & \textbf{0.73}\symnone \\[2.5pt]
\textbf{P-HF-2} & 5.52\symdag & 32.7\symdag & 0.23\symdag & \textbf{0.73}\symnone \\[2.5pt]
\textbf{P-HF-3} & \textbf{5.48}\symnone & \textbf{31.6}\symnone & \textbf{0.22}\symnone & 0.71\symnone \\[2.5pt]
\bottomrule[1.5pt]
\end{tabular}}
\vspace{2pt}
\begin{flushleft}
\footnotesize $^{\dagger}$ indicates a statistically significant difference compared with P-HF-3 (\( \text{p} < 0.05 \)).
\end{flushleft}
\end{table}

\begin{table}[!t]\large
\centering
\caption{Objective evaluation results based on Pseudo-patient-2 dataset.}
\label{tab:table4}
\resizebox{\linewidth}{!}{
\begin{tabular}{c|c|c|c|c}
\toprule[1.5pt]
\multirow{2}{*}{\textbf{Systems}} & \multicolumn{4}{c}{\textbf{Objective Evaluation Metrics}} \\ 
\cmidrule(lr){2-5}
 & \textbf{MCD} ($\downarrow$) & \textbf{CER} ($\downarrow$) & \textbf{F0 RMSE} ($\downarrow$) & \textbf{F0 CORR} ($\uparrow$) \\ 
\midrule
\textbf{Baseline 1} & 7.02\symdag & 53.7\symdag & 0.39\symdag & 0.78\symdag \\[2.5pt]
\textbf{Baseline 2} & 6.67\symdag & 44.9\symdag & 0.38\symdag & 0.80\symdag \\[2.5pt]
\addlinespace[1pt] \cline{1-5} \addlinespace[2.5pt]
\textbf{P-MF-1} & 6.63\symdag & 47.2\symdag & 0.37\symnone & \textbf{0.83}\symnone \\[2.5pt]
\textbf{P-MF-2} & 6.45\symdag & 43.5\symdag & \textbf{0.36}\symnone & 0.80\symdag \\[2.5pt]
\textbf{P-MF-3} & 6.44\symdag & 41.9\symdag & 0.37\symnone & 0.81\symnone \\[2.5pt]
\addlinespace[1pt] \cline{1-5} \addlinespace[2.5pt]
\textbf{P-IF-1} & 6.53\symdag & 44.6\symdag & \textbf{0.36}\symnone & \textbf{0.83}\symnone \\[2.5pt]
\textbf{P-IF-2} & 6.35\symdag & 40.8\symdag & 0.37\symnone & 0.81\symnone \\[2.5pt]
\textbf{P-IF-3} & 6.35\symdag & 39.4\symdag & \textbf{0.36}\symnone & \textbf{0.83}\symnone \\[2.5pt]
\addlinespace[1pt] \cline{1-5} \addlinespace[2.5pt]
\textbf{P-HF-1} & 6.48\symdag & 44.4\symdag & 0.38\symdag & 0.82\symnone \\[2.5pt]
\textbf{P-HF-2} & 6.31\symnone & 39.7\symdag & 0.37\symnone & 0.81\symdag \\[2.5pt]
\textbf{P-HF-3} & \textbf{6.29}\symnone & \textbf{38.2}\symnone & 0.37\symnone & 0.82\symnone \\[2.5pt]
\bottomrule[1.5pt]
\end{tabular}}
\vspace{2pt}
\begin{flushleft}
\footnotesize $^{\dagger}$ indicates a statistically significant difference compared with P-HF-3 (\( \text{p} < 0.05 \)).
\end{flushleft}
\end{table}

\subsubsection{Comparison with baselines} 

We first compare Baseline 1 with the ``-1'' systems, which use the same amount of original EL2SP data. The ``-1'' systems generally outperform Baseline~1 across all four datasets. This trend is further supported by statistical analysis on Patient-1 and Patient-2 datasets (11/12 and 9/12 significant comparisons, respectively; \( \text{p} < 0.05 \)). We then compare the ``-1'' systems with Baseline 2. Owing to its two-stage fine-tuning scheme and the use of much larger-scale SD, Baseline 2 generally outperforms the “-1’’ systems, confirming the training benefits brought by the rich knowledge in SD. Furthermore, despite using the limited original data, some ``-1'' systems achieve performance comparable to or slightly better than Baseline 2, such as on some metrics of Patient-1 (MCD, F0 RMSE, F0 CORR) and Pseudo-patient-2 (MCD, CER, F0 RMSE, F0 CORR) datasets in Tables~\ref{tab:table1} and~\ref{tab:table4}, respectively. These findings highlight the robustness of the proposed methods and suggest that, although Baseline 2 attempts to mitigate SD-related imperfections through progressive fine-tuning, speech-only representations still impose an inherent performance bottleneck.

Next, we compare Baseline 2 with the ``-2'' and ``-3'' systems. The overall results of the proposed systems are better than those of Baseline 2 across all datasets, with statistically significant improvements observed in 22/24 and 17/24 pairwise comparisons on Patient-1 and Patient-2 datasets, respectively, between Baseline 2 and the ``-2''/``-3'' systems (\( \text{p} < 0.05 \)). These results demonstrate that, in addition to the benefits gained from increased data volume, incorporating text representations provides guidance for learning a more effective mapping.

In summary, these comparisons across different EL2SP datasets clearly demonstrate the advantages of the proposed speech--text representation learning framework over speech-representation-based methods.

\subsubsection{Comparison among the proposed systems}

\begin{figure}[!t]
    \centering
    \begin{minipage}[t]{0.492\textwidth}
        \centering
        \includegraphics[width=\linewidth]{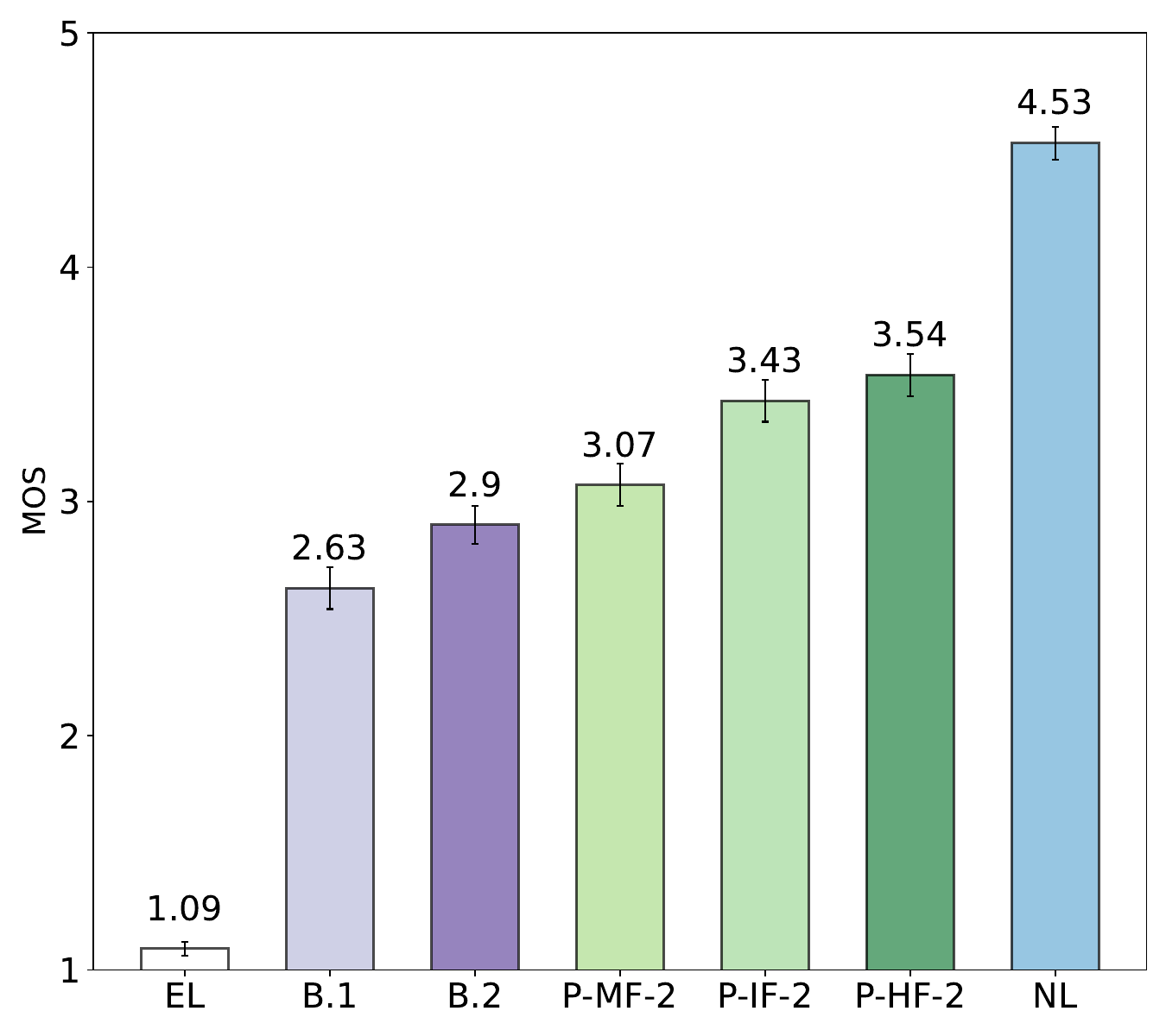}
        \centerline{\hfill(a) MOS for naturalness on Patient-1 dataset.\hfill}\medskip
    \end{minipage}
    \hfill
    \begin{minipage}[t]{0.492\textwidth}
        \centering
        \includegraphics[width=\linewidth]{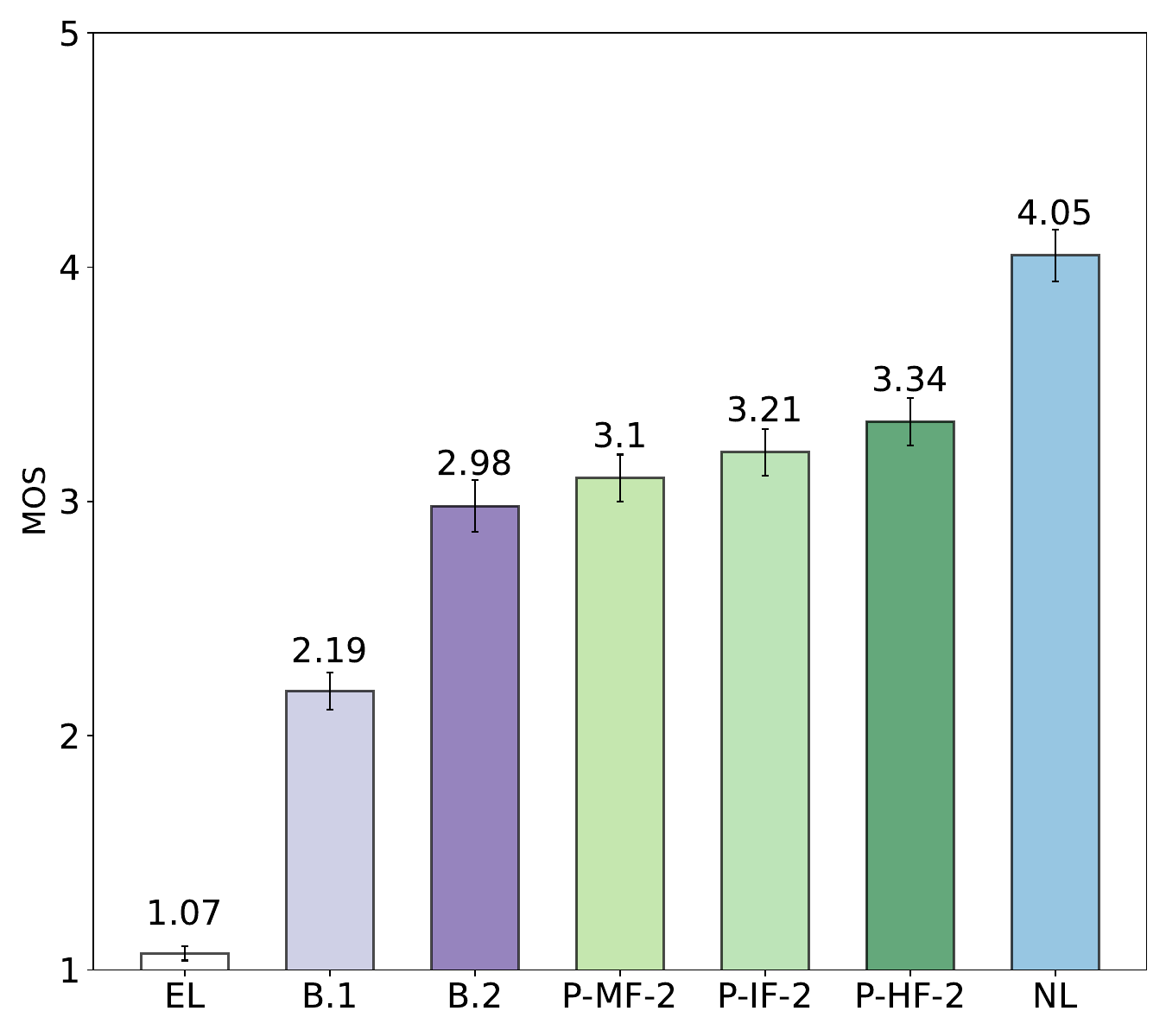}
        \centerline{\hfill(b) MOS for naturalness on Patient-2 dataset.\hfill}\medskip
    \end{minipage}
    \caption{MOS results with 95\% confidence intervals for naturalness. EL, NL, B.1, B.2 denote EL speech, target normal speech, and the converted speech from Baselines 1 and 2, respectively.}
    \label{fig:mos_test}    
\end{figure}

We focus on analyzing the performance of the proposed systems within the same fusion strategies, i.e., P-MF-1 vs. -2 vs. -3, P-IF-1 vs. -2 vs. -3, and P-HF-1 vs. -2 vs. -3. A generally improving trend emerges across all these fusion strategies. The ``-2'' systems incorporating SD yield substantial improvements over their ``-1'' counterparts (with statistical significance in 19/24 comparisons on Patient-1 and Patient-2 datasets, \( \text{p} < 0.05 \)), while the ``-3'' systems combining SD with the extended loss generally achieve the best results with relatively modest gains. These findings underscore the effectiveness of incorporating SD and the integrated-representation-guided loss for the proposed methods.

When examining systems that differ only in their fusion strategy, the proposed systems based on input-level fusion generally outperform their counterparts with middle-level fusion (P-MF-1 vs. P-IF-1, P-MF-2 vs. P-IF-2, and P-MF-3 vs. P-IF-3). This aligns with the design intuition, where input-level fusion assists the speech encoder in more directly accessing the text-enriched integrated representations during joint training. Consequently, the encoder can internalize text-related cues that remain beneficial later in the reconstruction stage with only speech inputs available. We also see a further improvement from input- to hybrid-level fusion strategies (P-IF-1 vs. P-HF-1, P-IF-2 vs. P-HF-2, and P-IF-3 vs. P-HF-3). By combining the structural advantages of middle- and input-level fusion, hybrid-level fusion facilitates a more refined integration of speech--text representations and strengthens the encoder's ability on reconstruction training.

In addition to these trends, we observe small fluctuations in pitch-related metrics, particularly F0 CORR. The statistical significance annotations also suggest significance is generally clearer for MCD and CER. These behaviors are expected for two reasons. First, imperfect SD inevitably carries prosodic inconsistencies that introduce instability into pitch modeling. Second, because EL2SP is essentially a speech synthesis task, incorporating text representations promotes modeling linguistic information, but causes small trade-offs in pitch reconstruction in converted speech. A similar phenomenon is noted in a previous work~\cite{leter2024strong}. 

Among all the proposed systems, P-HF-3 achieves the best performance in 16 out of the 20 comparable results across all five datasets. In addition, the statistical significance analysis provides further support for the overall superiority of P-HF-3. In particular, it outperforms the other systems in 31/40, 28/40, 34/40, 29/40, and 26/40 pairwise comparisons with statistical significance (\( \text{p} < 0.05 \)) on Patient-1, Patient-2, Patient-3, Pseudo-patient-1, and Pseudo-patient-2 datasets, respectively. Taken together, these findings reinforce the effectiveness of leveraging SD-based data augmentation, hybrid-level fusion, and the extended loss to enhance EL2SP performance.

\begin{figure}[!t]
    \centering
    \begin{minipage}[t]{\linewidth}
        \centering
        \includegraphics[width=\linewidth]{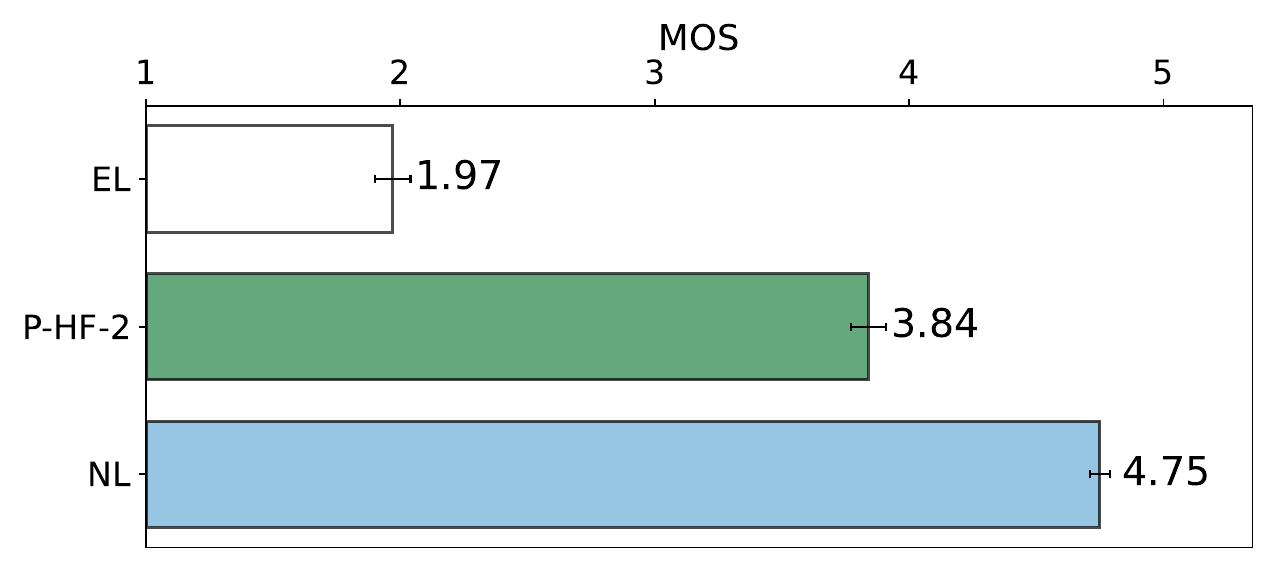}
        \centerline{\hfill(a) MOS for intelligibility on Patient-1 dataset.\hfill}\medskip
    \end{minipage}
    \hfill
    \begin{minipage}[t]{\linewidth}
        \centering
        \includegraphics[width=\linewidth]{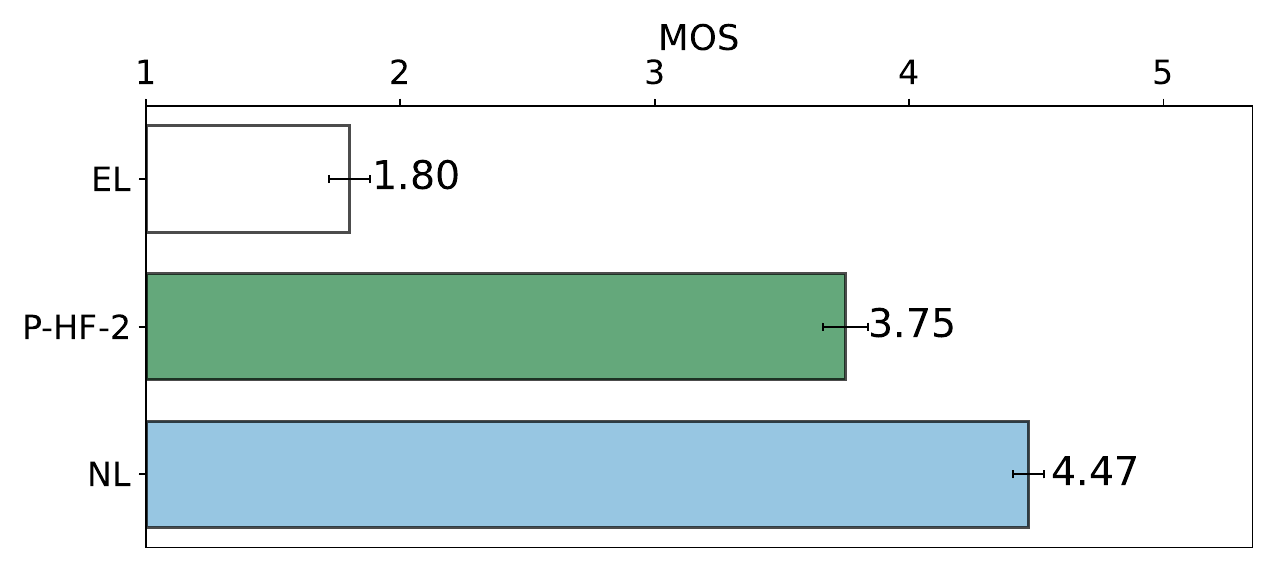}
        \centerline{\hfill(b) MOS for intelligibility on Patient-2 dataset.\hfill}\medskip
    \end{minipage}
    \caption{MOS results with 95\% confidence intervals for intelligibility. EL, NL denote EL speech and target normal speech, respectively.}
    \label{fig:mos_test2}    
\end{figure}

\subsection{Subjective Evaluation Results}

The subjective evaluations were conducted on Patient-1 and Patient-2 datasets to assess naturalness and intelligibility separately.

For the naturalness evaluation, the outputs of Baselines 1 and 2, P-MF-2, P-IF-2, and P-HF-2, along with their original EL and normal counterparts, were randomly selected for MOS tests. The obtained results are shown in Fig.~\ref{fig:mos_test}. As illustrated in Figs.~\ref{fig:mos_test}(a) and (b), all comparable systems yield significantly more natural speech than original EL speech under both datasets. Furthermore, all the proposed systems consistently outperform both baselines, indicating that the proposed methods of learning integrated speech--text representations leads to higher perceptual naturalness. Among the proposed systems, performance improves progressively from P-MF-2 onward, and P-HF-2 achieve higher MOS scores than P-IF-2 and P-MF-2 (for P-HF-2 compared with P-IF-2: \( \text{p} = 0.0277 \) for Patient-1 and \( \text{p} = 0.0141 \) for Patient-2; for P-HF-2 compared with P-MF-2: \( \text{p} < 10^{-15} \) for Patient-1 and \( \text{p} = 0.0003 \) for Patient-2), achieving closest naturalness to normal speech. These findings are similar to the objective evaluation trends, reinforcing the effectiveness of the proposed framework.

For the intelligibility evaluation, we used P-HF-2 as a representative proposed system. The original EL speech and target normal speech were also included as the lower and upper bounds, respectively. As illustrated in Figs.~\ref{fig:mos_test2}(a) and (b), the original EL speech is poorly intelligible, while P-HF-2 substantially improves subjective intelligibility over the original EL speech on both datasets, with mean MOS scores of 3.84 versus 1.97 on Patient-1 and 3.75 versus 1.80 on Patient-2 (both \( \text{p} < 10^{-15} \)). These observations are also consistent with the objective CER results.

Taken together, these results provide complementary subjective evidence that the proposed framework effectively improves not only naturalness but also intelligibility.

\subsection{Spectrogram Analysis}

\begin{figure*}[!t]
\centering{\includegraphics[width=1\linewidth]{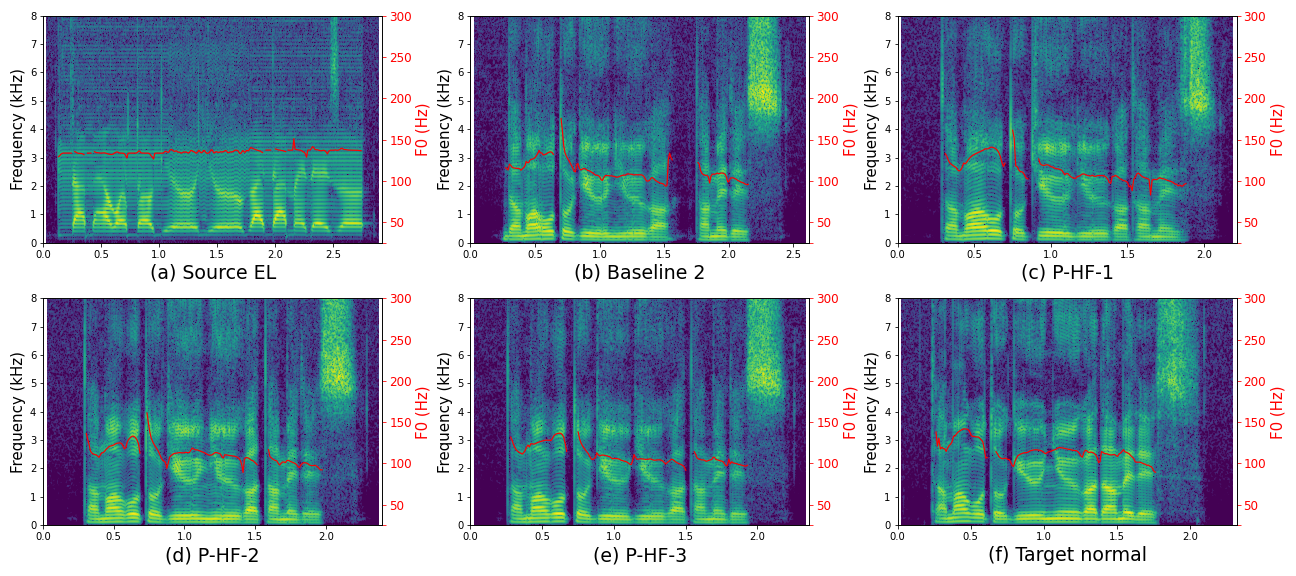}}
\caption{Spectrograms and F0 contour plots for a converted example from Baseline 2, P-HF-1, P-HF-2, and P-HF-3 on Patient-1 dataset. The horizontal axis represents time (seconds), and the vertical axis denotes frequency for both the spectrogram and the F0 contour.}
\label{fig6}
\end{figure*}

In Fig.~\ref{fig6}, we visualize the spectrograms and F0 contours of samples from Baseline 2, P-HF-1, P-HF-2, and P-HF-3, with EL and normal speech serving as lower and upper bounds, respectively. We have the following observations. As expected, the converted samples from the four systems exhibit much richer spectral patterns and more natural F0 contours than the EL speech, with temporal structures that more closely resemble the target normal speech. Furthermore, the proposed systems optimize the temporal structure compared with Baseline 2. P-HF-1 produces a more stable and coherent F0 contour than Baseline 2, despite relying solely on the limited original data. This aligns with the objective results shown in Table~\ref{tab:table1}, where Baseline 2 achieves better intelligibility, but underperforms P-HF-1 regarding speech-quality metrics such as MCD and F0 RMSE. P-HF-2 retains a duration comparable to P-HF-1 but provides more accurate spectral details, particularly in the ending region, where the spectral shape and energy distribution more closely match those of the target normal speech. Finally, P-HF-3 further enhances both the spectral structure and the progression of the F0 contour, yielding converted speech most closely approximating natural pronunciation. These findings are consistent with the quantitative trends observed in the objective evaluations.

\section{Discussion}

\subsection{Why Speech--Text Representations Are Effective}

The experimental results consistently emphasize the superiority of integrated speech--text representations over speech-only representations in EL2SP. The reason can be understood from two perspectives. First, EL speech exhibits severe acoustic mismatch relative to normal speech, including mechanical noise, abnormal excitation, reduced prosodic variability, and temporal inconsistency, as also illustrated in Fig.~\ref{fig2} and Fig.~\ref{fig6}. Under such conditions, relying only on speech input makes the linguistic content less transparent to the encoder. This makes it difficult to form stable and linguistically oriented intermediate representations, and may lead to cumulative mapping errors toward normal speech features. This also explains the still limited performance of the baseline systems relying only on speech representations. Second, compared with speech-only features, the corresponding text representations provide clearer, more structured, and speaker-independent linguistic cues during training, while remaining unaffected by EL-specific acoustic distortions. In addition, text inputs provide more stable cues for content and pause structure than EL speech features. This can also facilitate mappings toward normal temporal organization. Therefore, incorporating text guides the joint network toward more accurate intermediate representations and a more effective mapping to target normal speech. Furthermore, through the proposed reconstruction training, the benefits of these integrated speech--text representations are transferred to the final EL2SP model, thereby improving performance without requiring text input during inference.

\subsection{Sensitivity to Key Parameters}

To further clarify the effects of key parameters in the proposed method, we conducted a sensitivity analysis on the auxiliary loss weight $\lambda$ in Eq.~(14) and the number of attention heads in the MHA-based fusion module. The experiments were performed on Patient-1 dataset using P-MF-3 as the base configuration. Specifically, we evaluated 1) $\lambda \in \{0, 0.001, 0.01, 0.05\}$ while fixing the number of attention heads at 4, and 2) the number of attention heads in $\{2, 4, 8\}$ with $\lambda$ fixed at 0.01. The corresponding results are documented in Table~\ref{tab:sensitivity}. Here, $\lambda \ = 0.01$ with 4 heads corresponds to the initial P-MF-3 setting.

When varying $\lambda$, the results at $\lambda =0.001$ are close to those at $\lambda =0$ (i.e., P-MF-2, without using the additional loss term), with even slightly worse CER. This suggests that an overly small weight may be insufficient to provide effective guidance from the integrated representations, while slightly perturbing the optimization process. The initial setting $\lambda =0.01$ achieves the best overall performance. When $\lambda$ is further increased to 0.05, the results are slightly better than those at $\lambda =0$ and $\lambda = 0.001$, but remain inferior to those at $\lambda =0.01$. This indicates that the auxiliary guidance term is beneficial, while an excessively large weight may weaken the balance between the main reconstruction objective and the auxiliary guidance in Eq.~(14).

When varying the number of heads, 4 heads achieve the best performance, followed by 2 heads and then 8 heads. This suggests that 4 heads provide a better trade-off between attention diversity and per-head representation capacity in fusion module. Fewer heads may limit the diversity of cross-modal alignment patterns, whereas more heads may reduce the representation capacity of each head.

The sensitivity analysis supports the reasonableness of the chosen parameter settings. Meanwhile, the differences across nearby settings are relatively modest, indicating robustness of the proposed method within the tested parameter range.

\begin{table}[!t]
\centering
\caption{Sensitivity analysis using P-MF-3 with respect to $\lambda$ and attention heads.}
\label{tab:sensitivity}
\begin{tabular*}{\linewidth}{@{\extracolsep{\fill}} c|c|c|c}
\toprule[1.5pt]
\textbf{Parameter} & \textbf{Setting} & \textbf{MCD} ($\downarrow$) & \textbf{CER} ($\downarrow$) \\
\midrule
\multirow{4}{*}{\makecell[c]{$\lambda$ \\ (heads fixed at 4)}}
& 0     & 6.04          & 22.1          \\[2.5pt]
& 0.001 & 6.03          & 23.1          \\[2.5pt]
& 0.01  & \textbf{6.00} & \textbf{20.7} \\[2.5pt]
& 0.05  & 6.02          & 21.6          \\[2.5pt]
\addlinespace[1pt] \cline{1-4} \addlinespace[2.5pt]
\multirow{3}{*}{\makecell[c]{Heads \\ ($\lambda$ fixed at 0.01)}}
& 2     & 6.01          & 22.1          \\[2.5pt]
& 4     & \textbf{6.00} & \textbf{20.7} \\[2.5pt]
& 8     & 6.04          & 22.2          \\[2.5pt]
\bottomrule[1.5pt]
\end{tabular*}
\end{table}

\subsection{Computational Efficiency and Practicality}

To improve the convenience of daily communication for laryngectomees, computational efficiency is also an important factor for practical EL2SP applications, in addition to conversion performance. Through the proposed method, the finalized EL2SP systems achieve improved performance without introducing additional modules during inference. Thus, they maintain a typical autoregressive seq2seq framework with approximately 30.4 million trainable parameters, as mentioned in Section III-C. Specifically, on a single GPU, the end-to-end inference time for a speech utterance of approximately 5 seconds was about 0.87 seconds for seq2seq conversion (real-time factor (RTF) $\approx$ 0.17), with an additional 0.04 seconds for vocoder synthesis (RTF $\approx$ 0.01), resulting in an overall processing time of approximately 0.91 seconds (RTF $\approx$ 0.18). This performance is acceptable for some offline or assistive-use scenarios, for example, through server-side inference with GPU resources, where the input EL speech is sent and processed on a backend and then returned to the user device. Following such a deployment paradigm, prior work~\cite{nishio2026voice} based on a similar model structure has demonstrated the feasibility of deployment on mobile platforms, such as smartphones and iPads, for assistive speech conversion for laryngectomees. This suggests the practical potential of this type of EL2SP framework.

Nonetheless, the above processing time does not directly represent interaction latency. The main reason is that the current network is non-causal, i.e., the conversion process can start only after observing the end of an input sequence. Therefore, to achieve low-latency conversion, one essential research direction is to implement a causal architecture that can process input speech incrementally~\cite{11434701,zhao2025study,liang2025binauralflow}.  In addition, other potential directions include lightweight model design and leveraging non-autoregressive generation, as explored in~\cite{kameoka2021fasts2s,hayashi2022investigation,kobayashi2023low}, which may further reduce the computational cost and processing time.

\section{Conclusion}

This paper proposes a novel representation learning framework for seq2seq EL2SP that integrates speech and text representations. It consists of two main parts: 1) a speech--text representation integration part; and 2) a reconstruction training part. The former facilitates a more effective mapping between EL and normal speech by incorporating text representations; while the latter enables the final model to inherit the integrated representations without increasing network complexity, shedding light on practical applications. Building upon this framework, we develop a set of strategies including middle-, input-, and hybrid-level fusion strategies, together with an extended reconstruction objective with integrated-representation guidance. 

A series of systems are designed to examine the contribution of each component across multiple EL2SP datasets. The systematic experiments demonstrate that the proposed framework improves seq2seq EL2SP performance, yielding substantial gains over conventional speech-only baselines. The improvements become more pronounced when SD is incorporated as data augmentation. Furthermore, our analyses clarify the roles of different strategies: input-level fusion enables the encoder to more explicitly access text-enriched representations and therefore delivers stronger improvements than middle-level fusion; hybrid-level fusion combines the strengths of both schemes, achieving the most refined integration; and the extended loss enhances mapping accuracy in the finalized model. These findings collectively highlight the effectiveness of the proposed speech--text representation learning paradigm and offer insights for designing more powerful integration frameworks. 

Future work will explore extending the methodology to additional modalities, including visual information, e.g., using facial videos to recognize emotion and convey it by accordingly controlling the converted speech; evaluating robustness in more complex real-world environments; and developing lightweight, real-time EL2SP models suitable for on-device clinical applications by incorporating causal architectural designs.

\bibliographystyle{IEEEtran}
\bibliography{refs}
\end{document}